\documentclass[%
 prfluids,aps,
 amsmath,amssymb,superscriptaddress
]{revtex4-2}
\usepackage[utf8]{inputenc}
\usepackage{blindtext}
\usepackage{amsmath}
\usepackage{graphicx}
\usepackage{dcolumn}
\usepackage{mathrsfs}
\usepackage{subfig, color, caption}
\usepackage{bm}
\usepackage[english]{babel}
\usepackage{enumitem}
\usepackage{hyperref}
\usepackage{lipsum}
 \usepackage{mathtools}

\usepackage[normalem]{ulem}
\definecolor{darkgreen}{rgb}{0.07, 0.58 , 0.07}

\begin{document}

\title{Spirographic motion in a vortex}


\author{Sumithra Reddy Yerasi}
\email{yerasisumitra@gmail.com}
\affiliation{Universit\'e C\^ote d'Azur, CNRS, LJAD, 06100 Nice, France}
\author{Rama Govindarajan}%
 \email{rama@icts.res.in}
\affiliation{International Centre for Theoretical Sciences, Tata Institute of Fundamental Research, Bengaluru 560089, India
}%

\author{Dario Vincenzi}
\email{dario.vincenzi@univ-cotedazur.fr}
\thanks{Also Associate, International Centre for Theoretical Sciences,
Tata Institute of Fundamental Research, Bangalore 560089, India}
\affiliation{Universit\'e C\^ote d'Azur, CNRS, LJAD, 06100 Nice, France}

\date{\today}

\begin{abstract}
Studies of particle motion in vortical flows have mainly focused on point-like particles, either inertial or self-propelled.
This approximation assumes that the velocity field that surrounds the particle is linear.
We consider an inertialess rigid dumbbell in a two-dimensional steady vortex. While the system remains analytically tractable, the particle experiences the nonlinearity of the surrounding velocity field.
By exploiting the rotational symmetry of the flow, we reduce the problem to that of a two-dimensional dynamical
system, whose fixed points and periodic orbits can be used to explain the motion of the dumbbell.
For all vortices in which the fluid angular velocity decreases with radial distance,
the center of mass of the dumbbell follows a spirographic trajectory around the vortex center. This results from a periodic oscillation in the radial direction combined with revolution around the center. The shape of the trajectory depends strongly
on the initial position and orientation of the dumbbell, but the dynamics is qualitatively the same irrespective of the form of the vortex.
If the fluid angular velocity is not monotonic, the spirographic motion is altered by the existence of transport barriers,
whose shape is now sensitive to the details of the vortex.

\end{abstract}

\maketitle

\section{\label{sec:level1}Introduction}

The dynamical behaviour of particles in a single two-dimensional vortex can be complex in spite of the simple spatial structure of the flow. 
Particles that are denser than the fluid are ejected from the core of the vortex
at a rate that depends on the radial distance from the vortex center
\cite{ll89,rm97,cas04}.
This phenomenon generates strong inhomogeneities and even spikes in the spatial distribution of particles \cite{rm97, gl04,rg15}, with important
consequences on the collision and coalescence processes in vortical flows \cite{drg17}.
Light particles and bubbles behave in the opposite manner: they get entrapped into the vortex and accumulate near its center \cite{rm93,rm97}. Ejection, entrapment, and strong spatial heterogeneity in vortical flows
are also observed for inertialess but self-propelled particles, such as bacteria, plankton, or artificial microswimmers \cite{sa16,tml14, wilczek,bbbms21}.
The dynamical regimes depend critically on the motility, shape, and deformability of the particles, as well as on the magnitude of
rotational diffusion and external stimuli \cite{tml14, wilczek,gcbm21}.

In these studies, the particles are small enough to be treated as point-like. The velocity field surrounding them can therefore be modelled as linear, and if the particles possess internal degrees of freedom, the evolution of such degrees of freedom is entirely controlled by the local velocity gradient. 
Here our interest is to go beyond the point-particle approximation and explore the dynamics of an extended object that can experience the nonlinear structure of a flow field.
This is in general a  difficult problem, since even modelling the interaction
of such an object with the fluid and deriving the equations of motion may be a great challenge.
We consider a system that is sufficiently simple to allow an analytical study: an inertialess rigid dumbbell, with the two beads small enough in size to be in a Stokes flow relative to the fluid.  
This model is adapted from polymer physics, where it has been widely used to describe rodlike macromolecules \cite{graham}. 

The motion of the dumbbell is studied in a general two-dimensional steady vortex. 
By exploiting the rotational symmetry of the flow, the problem is reduced to the study of a two-dimensional dynamical system
which describes the position of the center of mass of the dumbbell and its orientation with respect to the radial direction.
The analysis of the fixed points and the periodic orbits of this system yields a complete understanding of the dynamics of the dumbbell.
In particular, if $\ell$ is the length of the dumbbell, $r_c$ the radial distance of its center of mass from the centre of the vortex, and $\alpha$ its
orientation angle, we show that the quantity $(r_c/\ell)\exp(-2r_c^2/\ell^2)\cos\alpha$ is a constant of the motion irrespective of the form of the vortex.
This result has different implications depending on the variation of the fluid angular velocity with the radial distance.
For all vortices in which the fluid angular velocity decreases with the radial distance, the dynamics is qualitatively
the same and consists of a spirographic quasiperiodic motion around the vortex center (here `spirographic' is used in a qualitative sense; it is not proved
that the trajectories are roulettes \cite{besant}).  The amplitude and the center of the radial oscillation can be predicted analytically and are found to
depend strongly on the initial configuration of the dumbbell.
For vortices where the fluid angular velocity is not strictly monotonic, the existence of
an attracting set in the configuration space alters the spirographic dynamics in a way that is specific to the vortex.
The attracting set indeed generates a barrier to transport in physical space, which is visualized by considering
the long-time spatial distribution of an ensemble of dumbbells.

Section~\ref{sect:dumbbell} outlines the equations governing the motion of the dumbbell and describes the vortical flow.
Section~\ref{sect:spirographic} exemplifies the spirographic dynamics by considering a dumbbell in a steady Lamb--Oseen vortex. 
The study of the fixed points and the periodic orbits of the reduced two-dimensional system  is presented in Sect.~\ref{sect:analytical} for a generic two-dimensional steady vortex. The Rankine vortex and a two-dimensional version of the Sullivan 
vortex are used to illustrate the case of a non-monotonic fluid angular velocity.
A summary of the results and some concluding remarks are given in Sect.~\ref{sect:conclusions}. 

\section{Rigid dumbbell in vortex 
flow}
\label{sect:dumbbell}

We consider a rigid dumbbell with two identical beads immersed in a Newtonian fluid. 
The connector between the beads does not pose any resistance to the fluid and should only be regarded
as a geometric constraint that maintains a fixed separation $\ell$. Moreover, $\ell$ is
assumed to be sufficiently large for hydrodynamic interactions between the beads to be negligible.
The motion of the fluid is described by the velocity field $\boldsymbol{u}(\bm x,t)$, and the force of the fluid
on each bead is given by the Stokes drag with coefficient $\zeta$. 

Let $\bm{r}_i$ ($i=1,2$) be the position vector of the $i$-th bead. Under the above asssumptions, $\bm r_i$ satisfies
\begin{equation}\label{eqnofmotion}
m \ddot{\bm{r}}_i = -\zeta [\dot{\bm{r}}_i - \bm{u} (\bm{r}_i, t)] + \bm{\tau}_i,
\quad i=1,2,
\end{equation}
where $m$ is the mass of each bead and $\bm{\tau}_i$ is the tension exerted by the connector on the $i$-th bead.
If the inertia of the beads is negligible, Eq.~\eqref{eqnofmotion}  simplifies to
\begin{equation}\label{velocity} 
\dot{\bm{r}}_i=\bm{u}(\bm{r}_i,t)+\frac{\bm{\tau}_i}{\zeta}, \quad i=1,2,
\end{equation}
and this study is conducted in the inertialess limit. The tension $\bm\tau_i$ can then be calculated by introducing the 
connector vector $\bm{\ell}=\bm{r}_1-\bm{r}_2$ and noting that the rigidity constraint can be written as
\begin{equation}\label{rigiditycons}
\bm{\ell}\cdot \dot{\bm{\ell}} = 0.
\end{equation}
We can thus subtract the equation for $\dot{\bm r}_2$ from that for $\dot{\bm r}_1$, 
take the dot product with $\bm \ell$, and equate the result to zero. Solving for $\tau_i=\vert\bm\tau_i\vert$ 
and observing that $\bm\tau_1=-\bm \tau_2$ is antiparallel to $\bm \ell$ then yields:
\begin{equation}\label{eq:tension}
\bm{\tau}_1 = -\bm\tau_2 =  - \frac{\zeta}{2}\,\{\hat{\bm{\ell}} \cdot [\bm{u}(\bm{r}_1,t)-
\bm{u}(\bm{r}_2,t)]\} \,  \hat{\bm{\ell}}
\end{equation}
with $\hat{\bm{\ell}}=\bm{\ell}/\ell$.
Equations~\eqref{velocity} and~\eqref{eq:tension} show that the motion of an inertialess dumbbell is independent of $\zeta$. 

As an alternative to the positions of the beads, the configuration of the dumbbell may be described by specifying the position of its center of mass,
$\bm r_c =(\bm r_1+\bm r_2)/2$, and the connector vector $\bm\ell$. The evolution equations for $\bm r_c$ and $\bm\ell$ are easily obtained from 
Eqs.~\eqref{velocity} and~\eqref{eq:tension}:
\begin{subequations}
\begin{eqnarray}
\dot{\bm r}_c &=& \dfrac{\bm u(\bm r_1,t)+\bm u(\bm r_2,t)}{2}, 
\label{eq:cm}
\\
\dot{\bm \ell} &=& \bm u(\bm r_1,t)-\bm u(\bm r_2,t) - 
\{\hat{\bm{\ell}} \cdot [\bm{u}(\bm{r}_1,t)- \bm{u}(\bm{r}_2,t)]\} \,  \hat{\bm{\ell}}.
\label{eq:ell}
\end{eqnarray}%
\label{eq:cm+ell}%
\end{subequations}
These equations generalize the rigid dumbbell model of polymer physics \cite{graham} to a nonlinear velocity field. Indeed, 
the usual polymer dumbbell model is obtained by replacing $\bm u(\bm x,t)=\bm u(\bm 0,t)+\nabla\bm u(t)\cdot\bm x$ into Eq.~\eqref{eq:cm+ell} 
(and adding Brownian fluctuations). An analogous generalization of the rigid dumbbell model was considered in
Ref.~\cite{pm09} in a study of gravitational settling in a cellular flow.

Here we focus on a steady two-dimensional vortex. In order to take advantage of the rotational symmetry of the flow, it is convenient to use the
polar coordinate system, where the position vector of a point with coordinates $(r,\varphi)$ is $\bm r=r(\cos\varphi,\sin\varphi)$ and
the unit vectors that form the orthogonal basis at the point $(r,\varphi)$ are $\hat{\bm r}=(\cos\varphi,\sin\varphi)$ and $\hat{\bm \varphi}=
(-\sin\varphi,\cos\varphi)$. We take a velocity field of the form
\begin{equation}\label{eq:velocity}
\bm u(\bm r) = U(r) \, \hat{\bm \varphi},
\end{equation}
where $U(r)$ is the azimuthal velocity. Therefore, the fluid angular velocity at a distance $r$ from the center of the vortex is 
\begin{equation}\label{eq:omega}
\Omega(r)=\frac{U(r)}{r}.
\end{equation}
In Sect.~\ref{sect:analytical}, we will show that several properties of the dynamics of the dumbbell can be 
predicted from Eqs.~\eqref{eq:cm+ell}.
The analytical study is
not confined to any specific choice of the function $\Omega(r)$ and holds for a general steady two-dimensional vortex flow.
However, to gain intuition on the dynamics, in the next section we first show the results of numerical simulations for
a two-dimensional, time-independent Lamb--Oseen vortex. As we shall see, the motion of the dumbbell in this
vortex is representative of the motion in any vortex such that $\Omega(r)$ decreases with $r$.

\section{Spirographic dynamics}
\label{sect:spirographic}

\begin{figure}[t]
\centering
\includegraphics[width=0.33\textwidth]{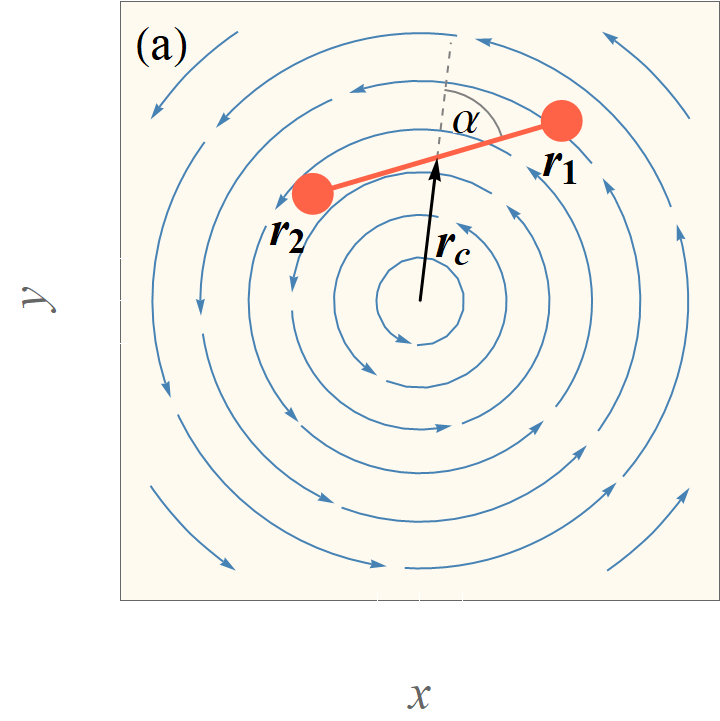}%
\hfill%
\includegraphics[width=0.33\textwidth]{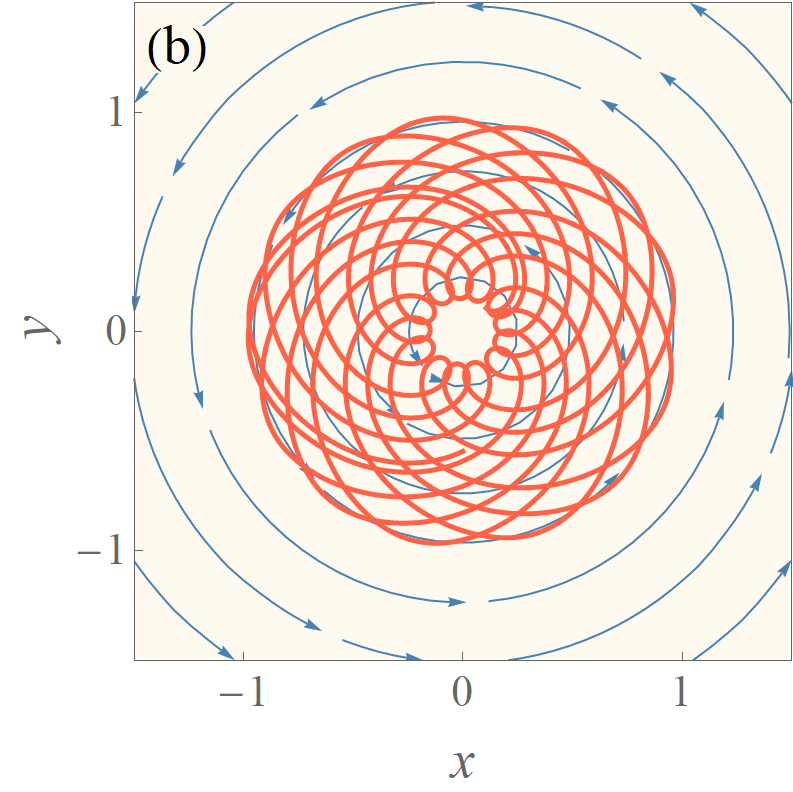}%
\hfill%
\includegraphics[width=0.33\textwidth]{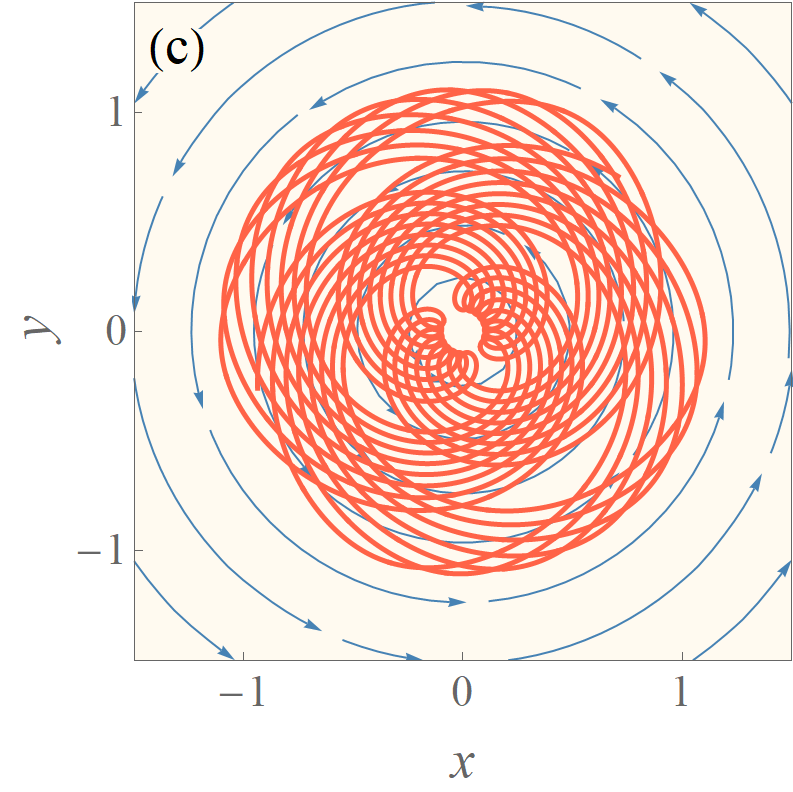}%
\caption{(a) Schematic of the dumbbell in a vortex. (b) Trajectory of the center of mass of the dumbbell in the Lamb--Oseen vortex for $r_c(0)=0.3$, $\varphi_c(0)=\pi/4$, $\alpha(0)=0$. 
(c)  The same as in (b) for $r_c(0)=1$, $\alpha(0)=-\pi/4$.}
\label{fig:spirographs}
\end{figure}

In the steady, two-dimensional Lamb--Oseen vortex, the
angular velocity is 
\begin{equation}
\Omega(r)=\frac{\Gamma}{2\pi}\,\frac{1-e^{-r^2/R^2}}{r^2},
\end{equation}
where $R$ is the size of the vortex core and $\Gamma$ its circulation.
Equations~\eqref{eq:cm+ell} are integrated by using
a second-order Heun method with time step $dt=10^{-4}$, which is sufficient to keep the length of the connector constant. Unless otherwise specified, the simulation parameters are $R=0.1$, $\Gamma=2\pi $, and $\ell=1$.

Figures~\ref{fig:spirographs}(b) and~\ref{fig:spirographs}(c) 
show two representative trajectories of the center of mass of the dumbbell (see also Supplemental Movies~1 \cite{fig1b} and~2 \cite{movie1c}).
This oscillates back and forth between two concentric circles while simultaneously revolving around the center of the vortex. The combination of 
these two motions generates a spirographic trajectory that eventually fills an annulus around the vortex center. 
The shape of the trajectory and the way it is covered are found to depend strongly
on the initial position and orientation of the dumbbell [compare Figs.~\ref{fig:spirographs}(b) and~\ref{fig:spirographs}(c)].

Because of the rigidity constraint, the dumbbell only possesses three degrees of freedom. 
It is therefore useful to describe its configuration
by means of the polar coordinates of the center of mass, $(r_c,\varphi_c)$, 
and the angle $\alpha$ that $\bm\ell$ makes with $\bm r_c$. 
This angle gives the
orientation of the dumbbell with respect to the radial direction [see Fig.~\ref{fig:spirographs}(a)];
$\alpha=0$ when the connector is parallel to the radial direction and it increases anticlockwise.
For reasons that will be clear later, it is convenient to take $-\pi/2\leqslant\alpha<3\pi/2$.
When $\alpha=0$ ($\alpha=\pi$) the dumbbell is parallel (antiparallel) to the radial direction; when $\alpha=\pm\pi/2$ the dumbbell is perpendicular to it and hence tangent to the streamlines of the vortex.
Note that the value of $\alpha$ also determines which of the beads is closest to the vortex center:
for $\pi/2<\alpha<3\pi/2$ bead ``1'' is closest to the centre, while 
for $-\pi/2<\alpha<\pi/2$ bead ``2'' is closest.

\begin{figure}[t]
\begin{minipage}[c]{.31\textwidth}
\includegraphics[width=\textwidth]{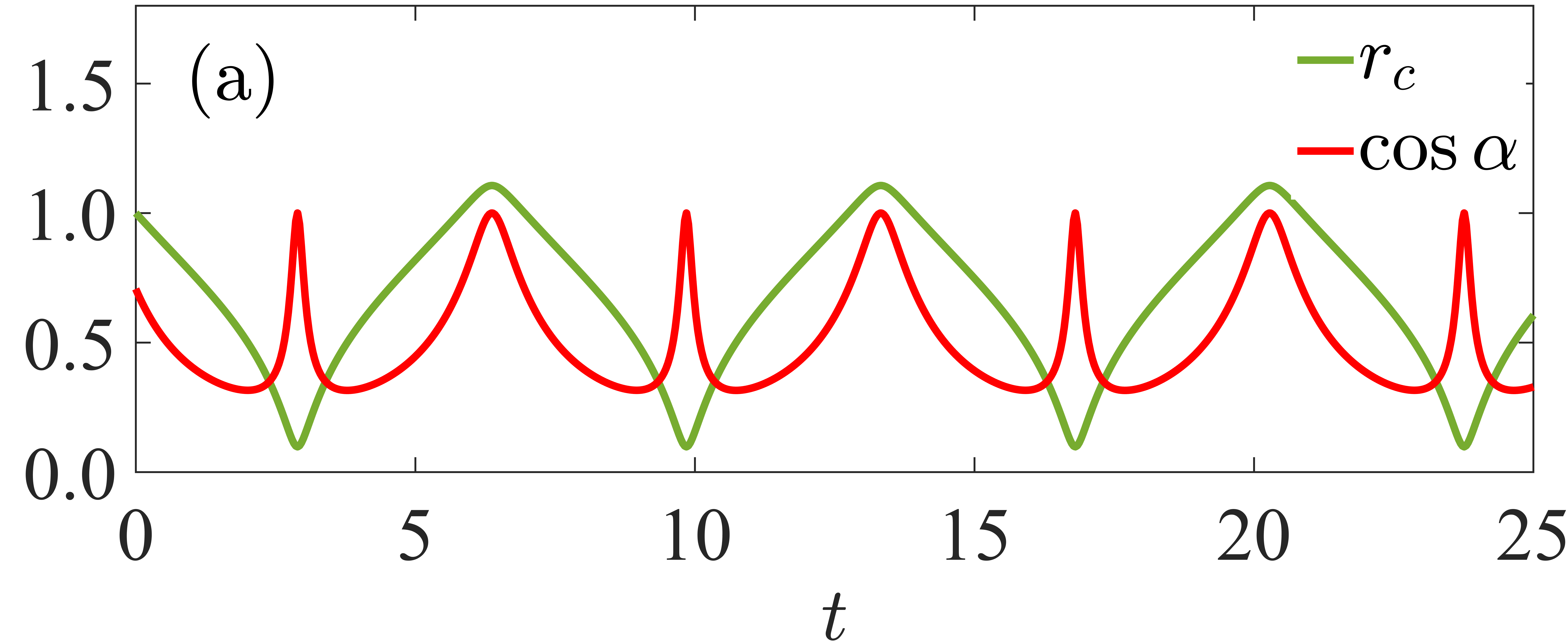}
  \vspace{4mm}
  \includegraphics[width=\textwidth]{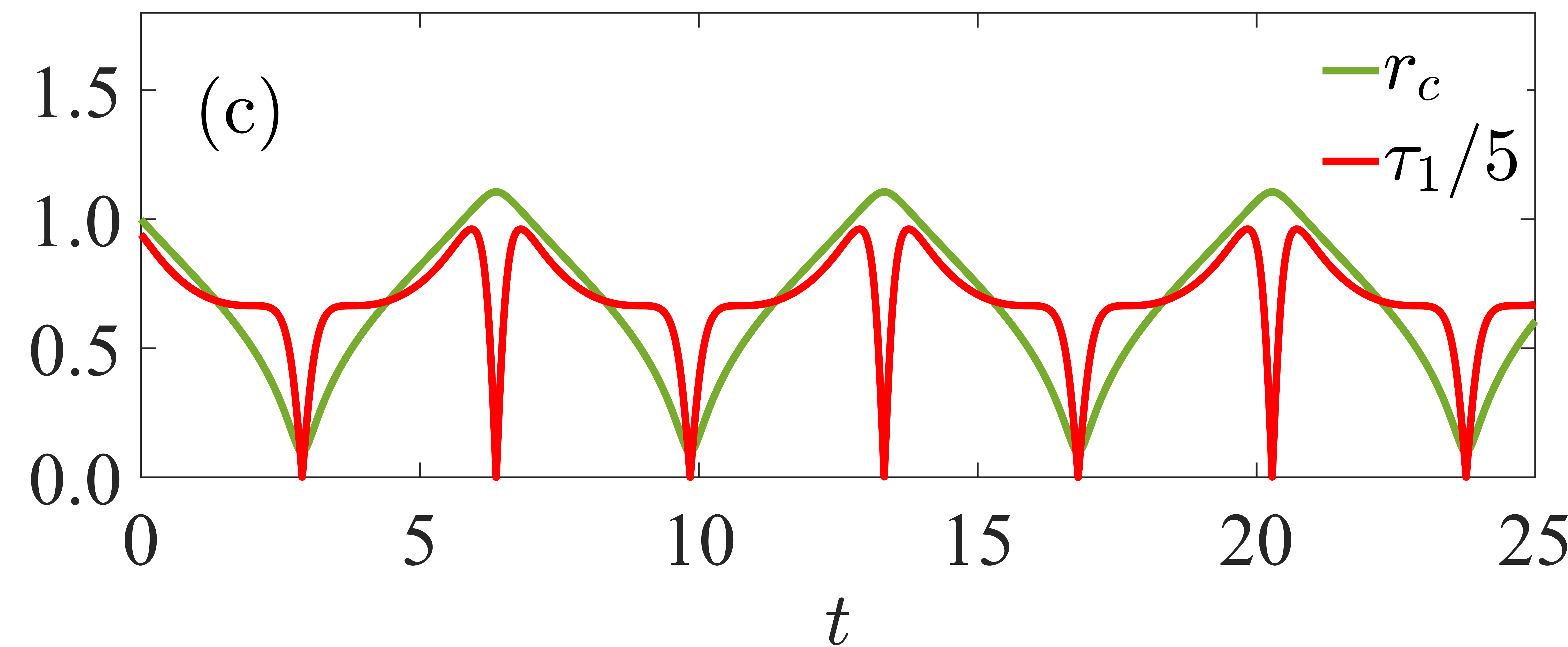}
\end{minipage}%
\hspace{1.2mm}
\begin{minipage}[c]{.31\textwidth}
  \includegraphics[width=1\textwidth]{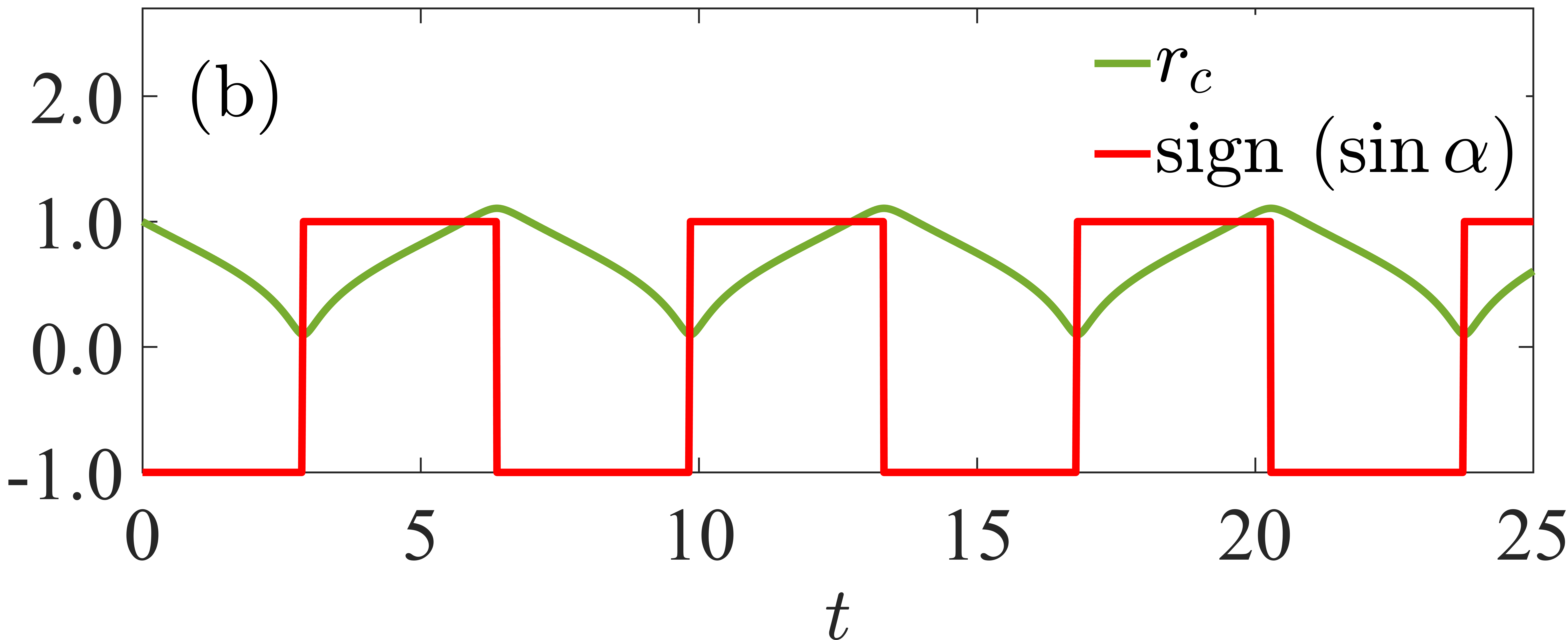}
  \vspace{4mm}
  \includegraphics[width=\textwidth]{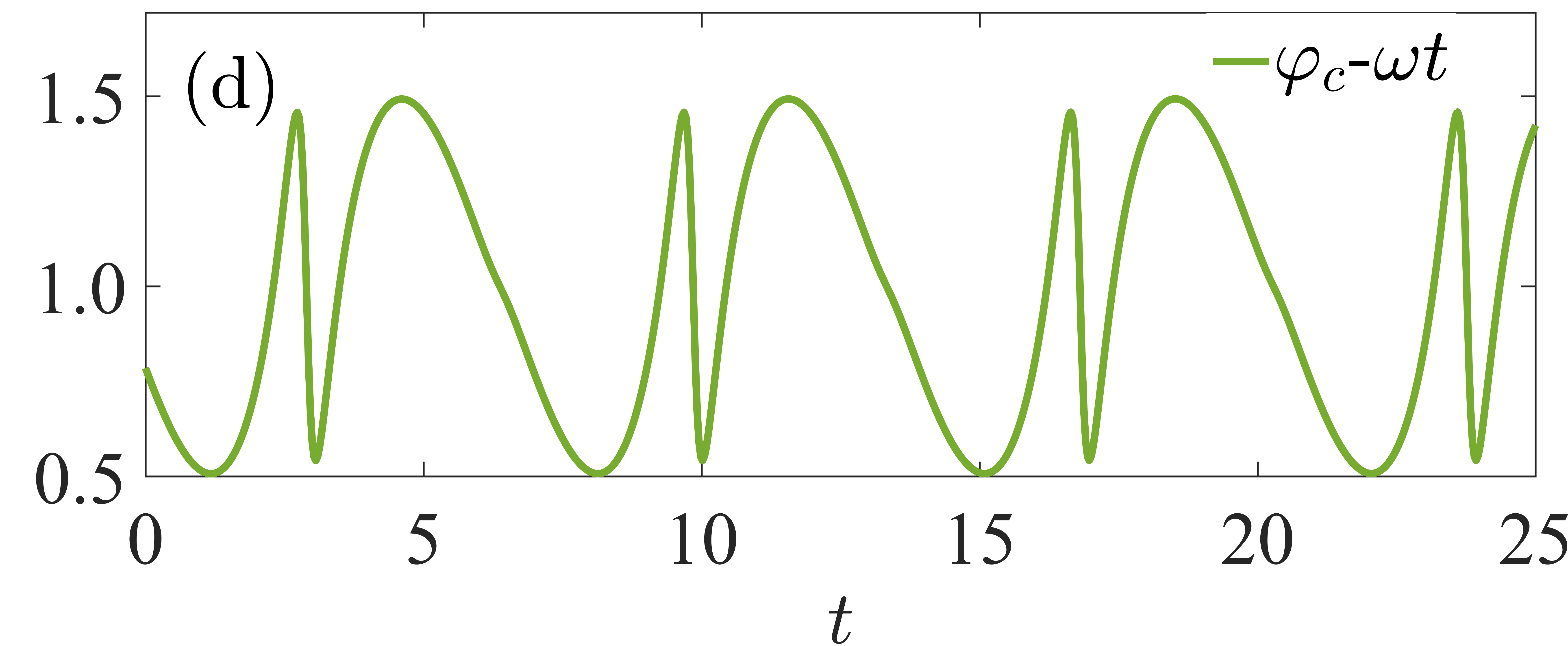}
\end{minipage}%
\hspace{1.2mm}
\begin{minipage}[c]{.33\textwidth}
  \includegraphics[width=\textwidth]{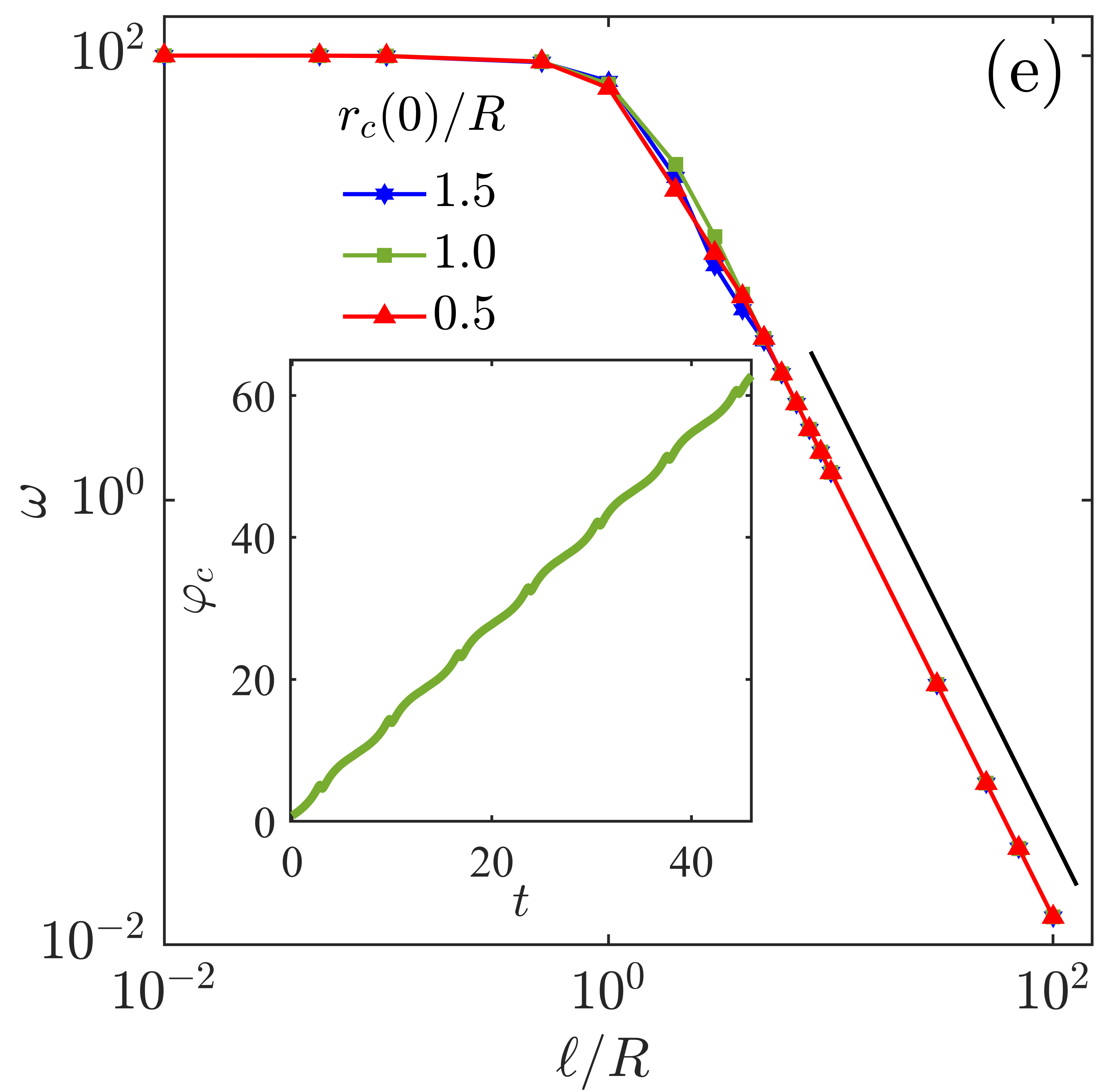}
\end{minipage}%
 \caption{Dumbbell in the vicinity of a Lamb--Oseen vortex: the time evolution of (a) $r_c$ and $\cos{\alpha}$, (b) $r_c$ and the sign of $\sin{\alpha}$, (c) $r_c$ and $\tau_1$, (d) $\varphi_c-\omega t$; in (c), the magnitude of $\tau_1$ is divided by 5 to make the comparison of the curves easier.
The initial conditions are the same
 as in Fig.~\ref{fig:spirographs}(c). Here $\omega=4.7$ and $T=6.9$. (e) Dependence of $\omega$ on $\ell/R$ for $\alpha(0)=-\pi/18$ and different values of $r_c(0)/R$. The black line is proportional to $(\ell/R)^{-2}$. The inset shows the time series of $\varphi_c$ for the same initial conditions as in panels (a)-(d).
 }
\label{timeseries}
\end{figure}

Representative time series of $r_c$, $\varphi_c$, $\cos\alpha$, and the tension in the connector 
are shown in Fig.~\ref{timeseries}; 
inspection of these time series clearly describes the dynamics of the dumbbell. 
Both $r_c$ and $\cos\alpha$ are periodic with same 
time period $T$ [Fig.~\ref{timeseries}(a)]. The distance of the dumbbell from the vortex centre 
oscillates between a minimum and a maximum value, so that the motion is confined to an annulus concentric with the vortex. 
The maximum and minimum distances are reached when $\cos\alpha=1$, {\it i.e.}~when the dumbell is parallel to the radial direction.
In such configuration, the tension in the connector vanishes [Fig.~\ref{timeseries}(c)].
Note that $\cos\alpha$ never changes sign. This means that, during the
motion, the connector vector keeps its initial, either inward or outward, orientation with respect to the radial direction
and never reverses it.
In other words, the bead that starts closest to the center of the vortex always remains closest to it (see also
Supplemental Movies~1 \cite{fig1b} and~2 \cite{movie1c}).
Finally, the evolution of $\varphi_c$ is the combination of a linear growth with slope $\omega$ (which corresponds to a
rotation about the vortex with angular velocity $\omega$) and a periodic oscillation with same time period as $r_c$ and $\cos \alpha$ [Figs.~\ref{timeseries}(d,e)].
Since $2\pi/\omega \neq k T$, where $k$ is a rational fraction, the angular motion is not periodic, and hence the trajectory of the center of mass never repeats itself but fills an annulus around the vortex center, in classic quasiperiodic motion. Fig.~\ref{timeseries}(e) suggests that $\omega$ is independent of the initial conditions when the ratio $\ell/R$ is
either very large or very small. In contrast, for intermediate values of $\ell/R$, 
$\omega$ depends on $r_c(0)$ and $\alpha(0)$.
Moreover, $\omega$ scales as $(\ell/R)^{-2}$ for $\ell/R\gg 1$, while it tends to a constant as $\ell/R\to 0$, \textit{i.e.}
the dynamics of the dumbbell does not reduce to that of a point particle in the $\ell/R\to 0$ limit. 

\begin{figure}[t]
\begin{minipage}[c]{.25\textwidth}
\includegraphics[width=\textwidth]{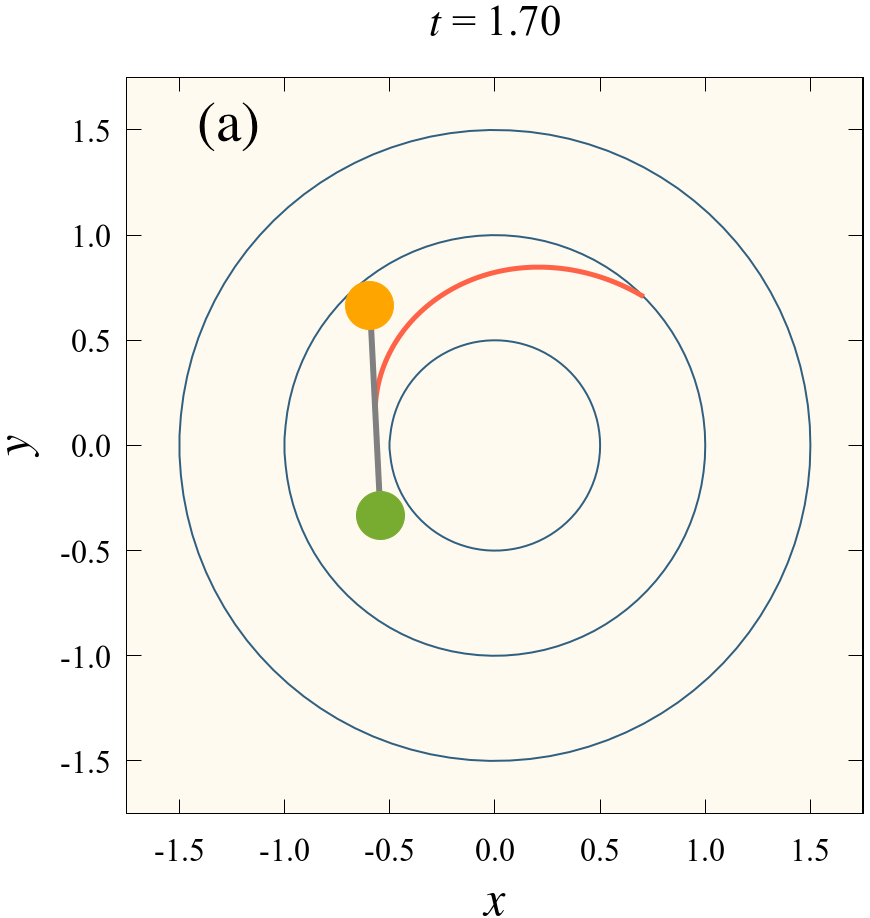}
\end{minipage}%
\hfill
\begin{minipage}[c]{.25\textwidth}
\includegraphics[width=1\textwidth]{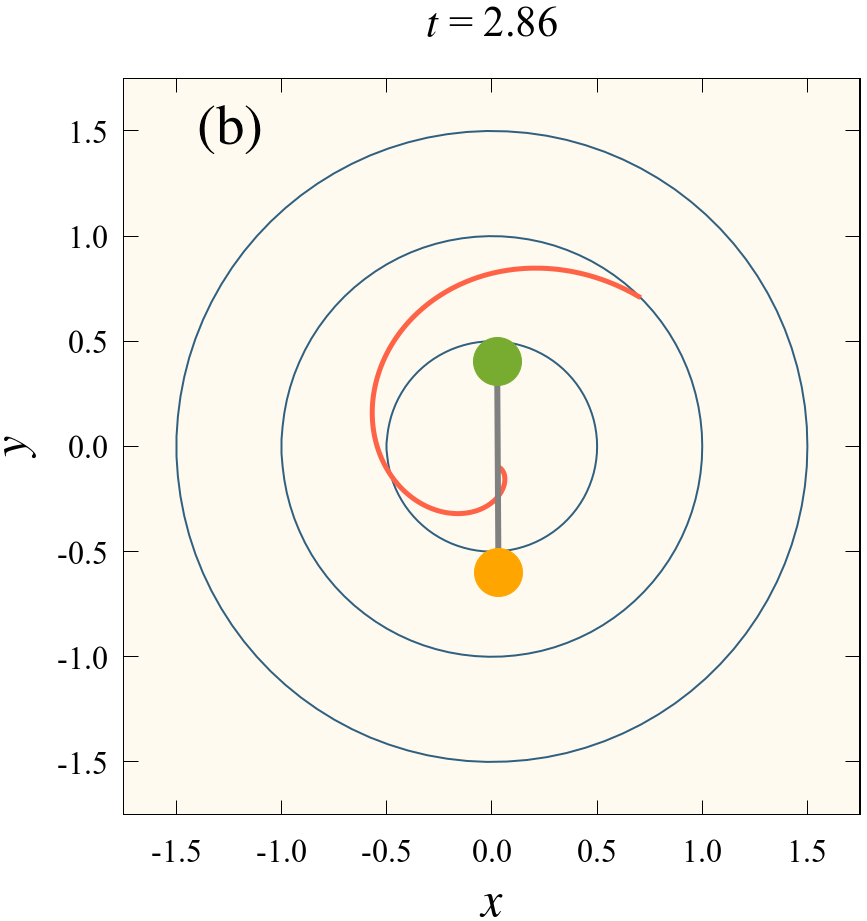}
\end{minipage}%
\hfill
\begin{minipage}[c]{.25\textwidth}
\includegraphics[width=\textwidth]{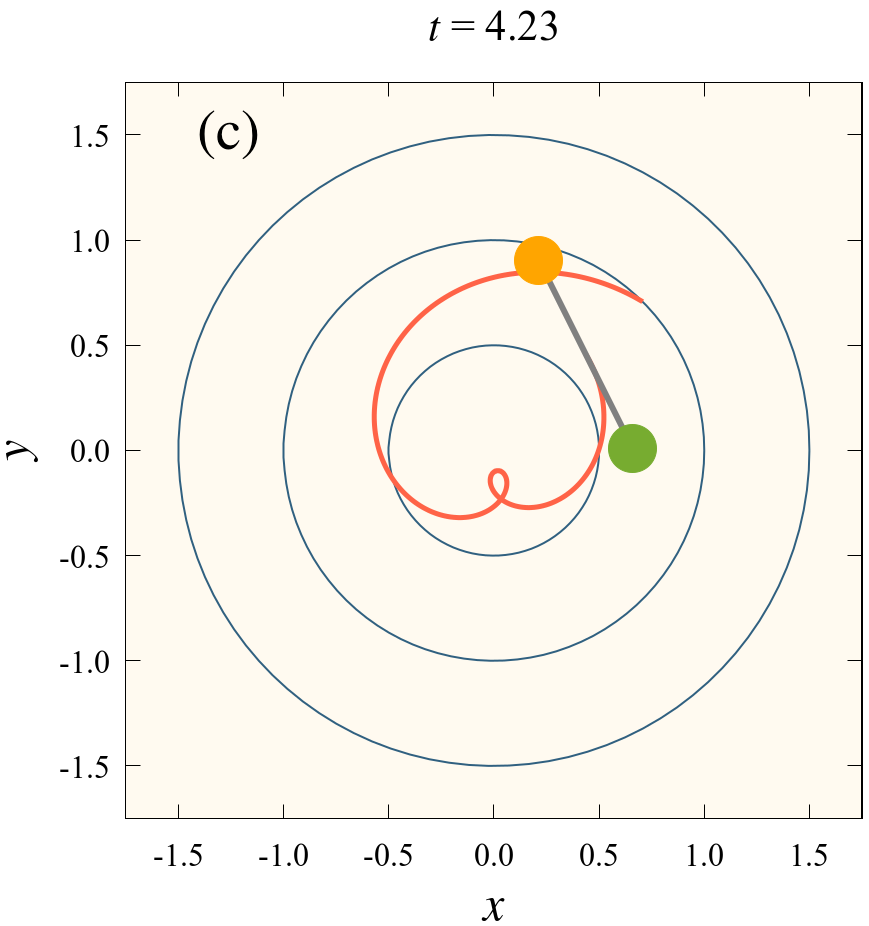}
\end{minipage}%
\hfill
\begin{minipage}[c]{.25\textwidth}
\includegraphics[width=\textwidth]{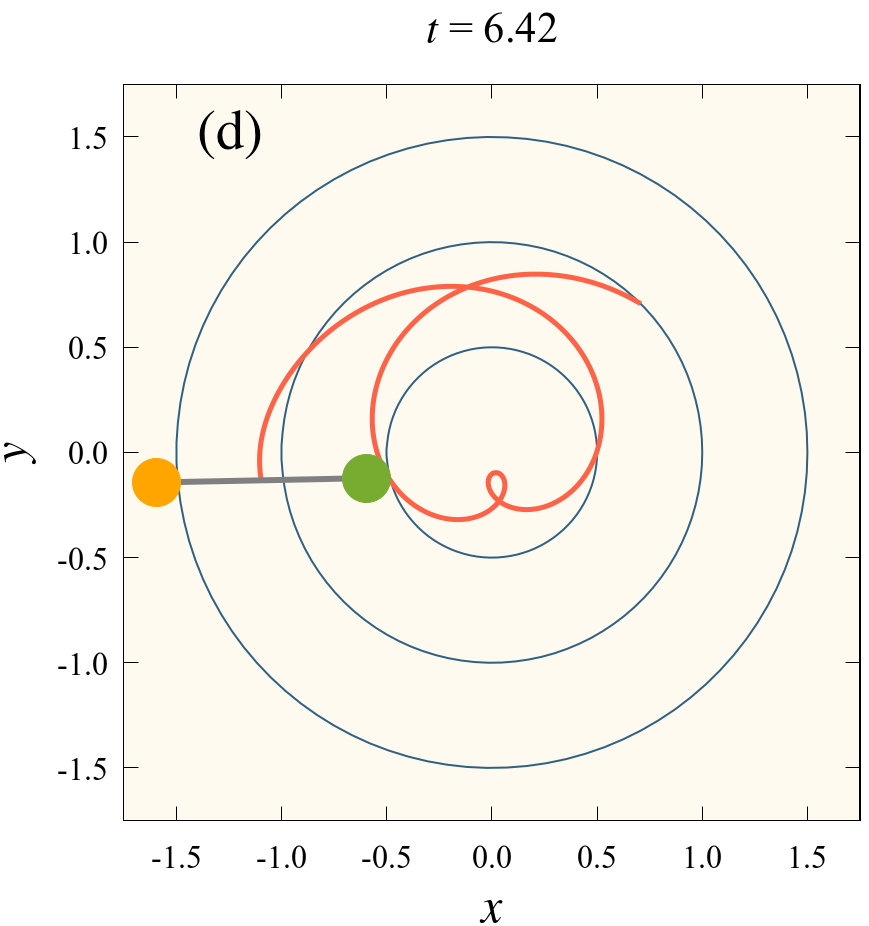}
\end{minipage}%
\caption{Position and orientation of the dumbbell in the vicinity of a Lamb--Oseen vortex at typical times. These snapshots correspond to the trajectory shown in Fig.~\ref{fig:spirographs}(c); see also Supplemental Movie~2 \cite{movie1c}.}
\label{fig:snapshots}
\end{figure}

In summary, the motion of the dumbbell can be described as the superposition of: 
i) a periodic oscillation with period $T$ of the center of mass in the radial direction; ii) a periodic revolution of the center of mass around the vortex center with a period $2\pi/\omega$, which is not in general a rational multiple of $T$;
iii) a periodic oscillation with period $T$ of the connector around the centre of mass of the dumbbell without reversals. In our simulations, we did not find any instance of periodic motion, but in principle
there may be some special values of $r_c(0)$ and $\alpha(0)$ such that $2\pi/\omega$ is a rational multiple of $T$, in which case the spirograph would not be space filling.
The resulting spirographic dynamics can also be  described as follows [see 
Figs.~\ref{timeseries}(b) and Fig.~\ref{fig:snapshots} as well as Supplemental Movies~1 \cite{fig1b} and~2 \cite{movie1c}].
Let us consider an initial configuration in which bead ``2'' is
closest to the vortex center ($-\pi/2 \leqslant \alpha(0)\leqslant \pi/2$) and hence has a higher angular velocity. 
When  bead ``2''
is ``leading'' 
($-\pi/2 < \alpha < 0$), the dumbbell moves inwards [Fig.~\ref{fig:snapshots}(a)], while its orientation gradually
approaches the radial direction ($\alpha$ increases). 
The inward motion continues until the dumbbell aligns with the radial direction ($\alpha=0$) [Fig.~\ref{fig:snapshots}(b)],
after which  bead ``2'' starts ``lagging'' ($0 < \alpha < \pi/2$) and the dumbbell moves outwards [Fig.~\ref{fig:snapshots}(c)].
Once the dumbbell aligns again with the radial direction, the inward motion restarts [Fig.~\ref{fig:snapshots}(d)].

Qualitatively the same dynamics to that shown in Figs.~\ref{timeseries} and~\ref{fig:snapshots}
is observed for different initial positions and orientations of the dumbbell as well as different values of the 
parameters $\ell$ and $R$. 
Because of the rotational symmetry of the problem, the dynamics of the dumbbell is obviously independent of $\varphi_c(0)$.
However, the details of the motion depend very sensitively on the other initial conditions
and on the system parameters. We demonstrate this by focusing on the time evolution
of the distance $r_c$. This can be described as
\begin{equation}
r_c(t)= r_c^\star + A f\left(\frac{t-t^\star}{T}\right),
\end{equation}
where $r^\star_c$ is the distance around which the oscillation takes place, $A$ its amplitude, $T$ the time period over which $\alpha$ and $r_c$ go through one cycle, and $t^\star$ a chosen temporal translation.
The function $f(z)$ is periodic of unit period and such that $-1\leqslant f(z)\leqslant 1$, $f(0)=1$, and $\int_0^1 f(z)dz=0$.

\begin{figure}[!t]
\begin{center}
\captionsetup[subfigure]{labelformat=empty,labelsep=none}
\subfloat[]
{\includegraphics[width=0.33\textwidth]{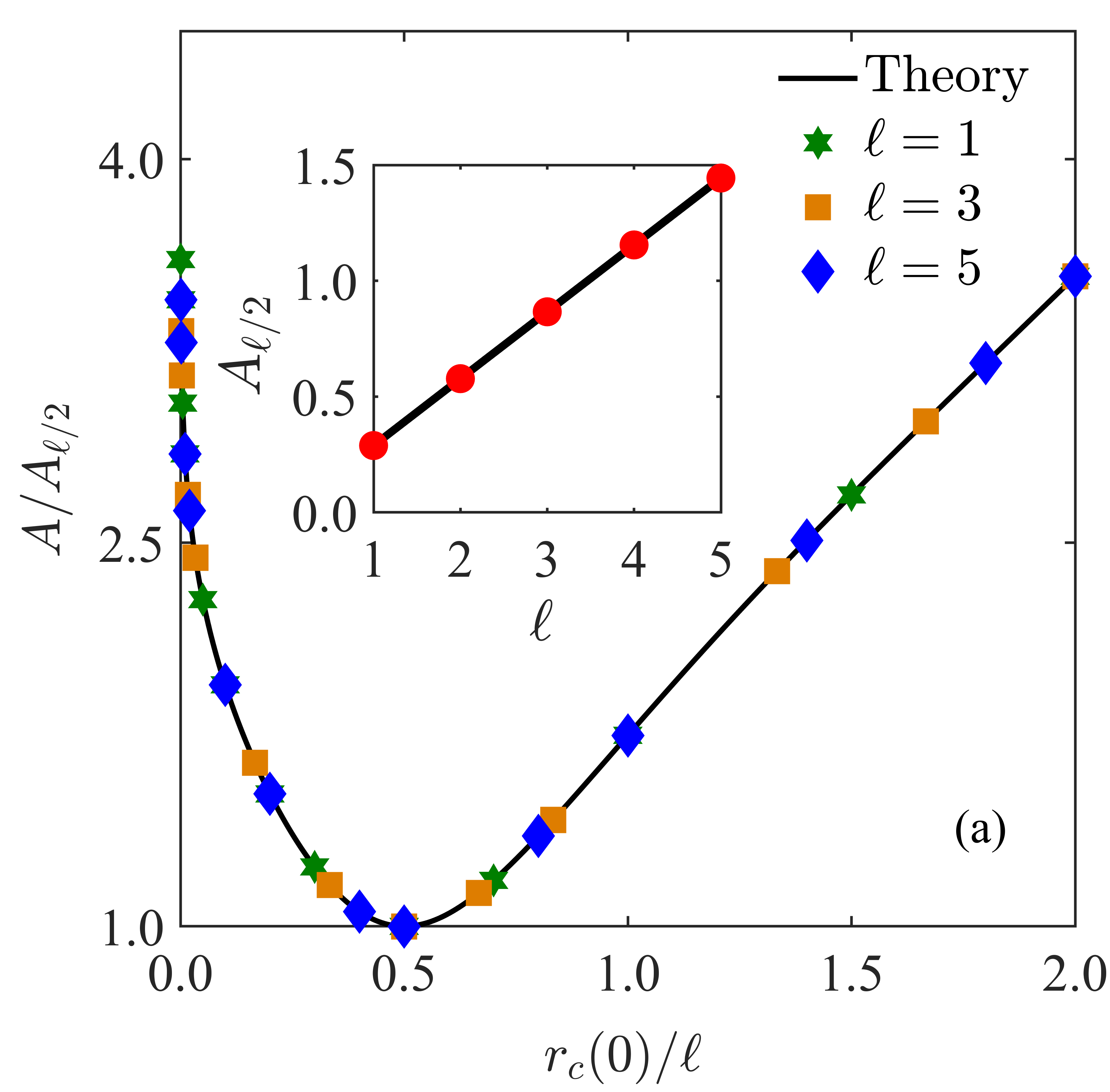}}
\subfloat[]
{\includegraphics[width=0.33\textwidth]{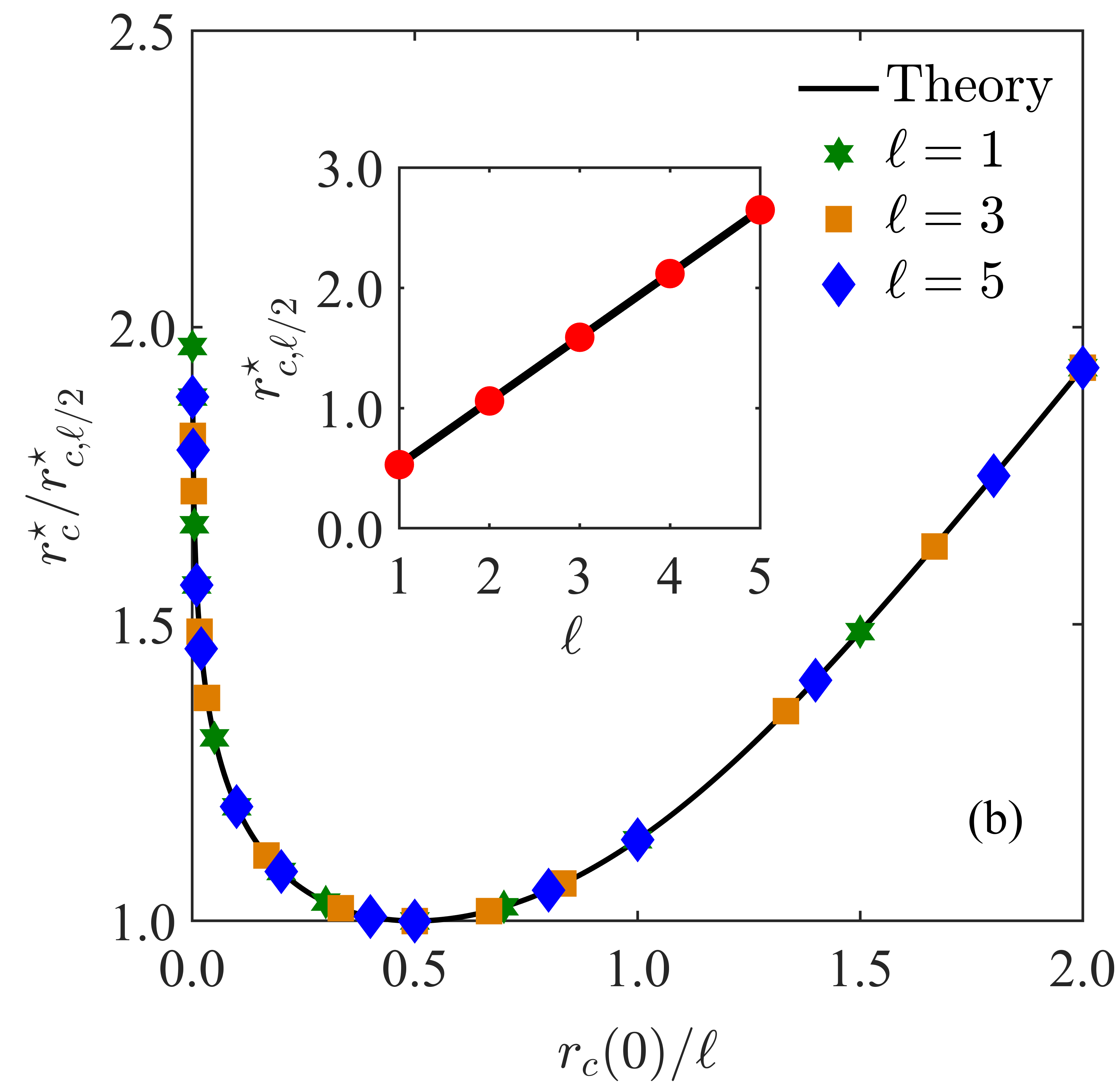}}
\subfloat[]{\includegraphics[width=0.33\textwidth]{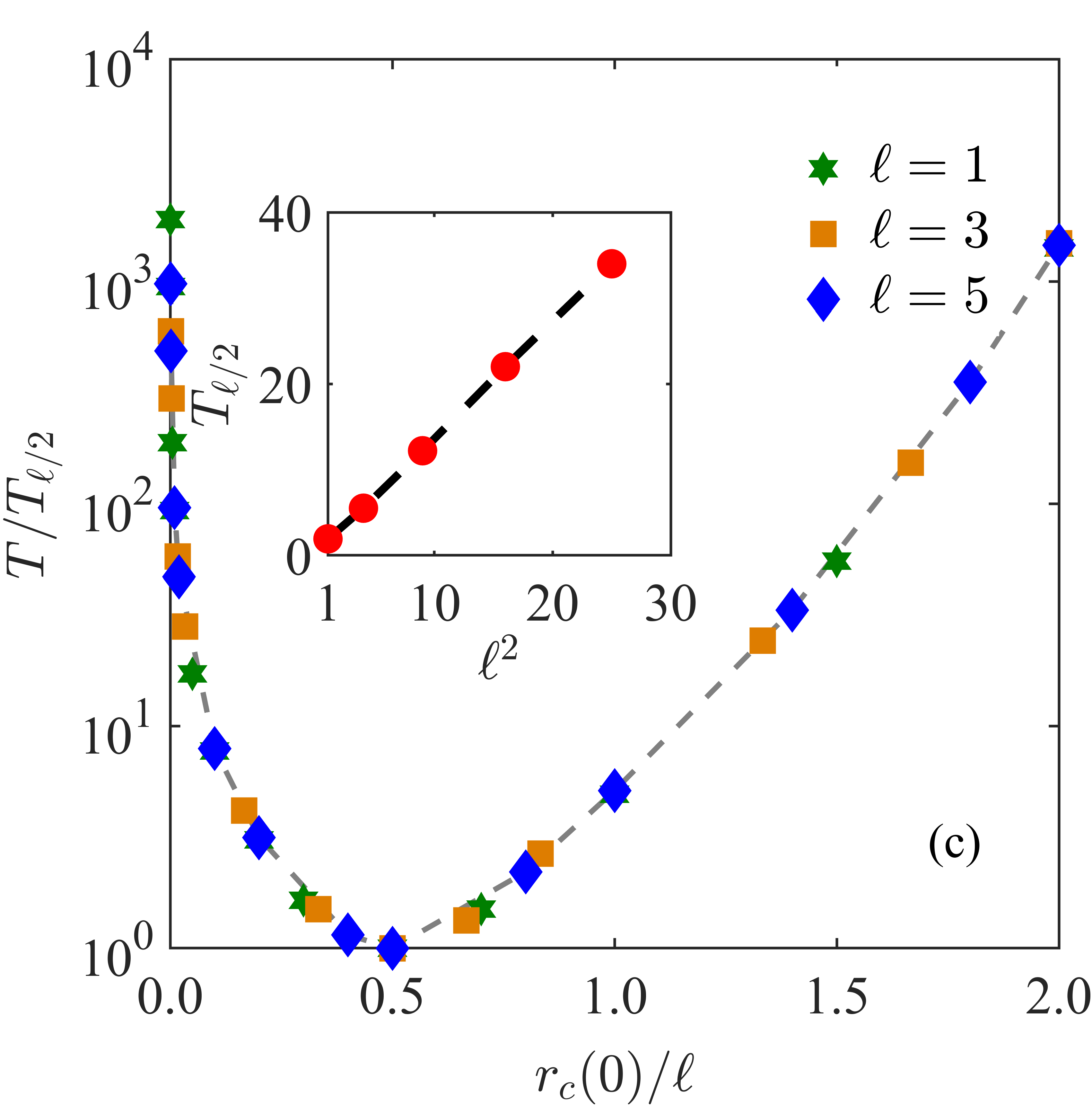}}
\caption{Motion of the dumbbell near a Lamb--Oseen vortex: dependence on the initial distance $r_c(0)$ of (a) the amplitude $A$, (b) the base radial distance $r_c^{\star}$ around which the center of mass oscillates, and (c) the time period $T$ for fixed $\alpha(0)=\pi/4$. 
$A_{\ell/2}$, $r_{c,\ell/2}^*$, $T_{\ell/2}$ are the values of  $A$, $r_{c}^*$, $T$ at $r_c(0)=\ell/2$ and $\alpha(0)=\pi/4$.
The insets show the dependence of these quantities on $\ell$. 
The solid lines are obtained from Eq.~\eqref{eq:A+rstar} (see Sect.~\ref{sect:analytical}); the dashed lines are included to guide the eye.}
\label{Artvsr0}
\end{center}
\end{figure}

For a fixed initial orientation $\alpha(0)\neq\pm\pi/2$, the quantities
$A$, $r^{\star}_c$, $T$ are convex functions of the initial distance $r_c(0)$;
they reach their minima when $r_c(0)=\ell/2$ and diverge as $r_c(0)$ approaches zero or becomes very large (see Fig.~\ref{Artvsr0}).
Thus, the oscillations performed by the center of mass are wider when the dumbbell is initially placed
at a distance either much smaller or much greater than half the length of the connector.
Rescaling $A$, $r_c^\star$, $T$ with their minimum values (denoted as $A_{\ell/2}$, $r_{c,\ell/2}^*$, $T_{\ell/2}$)
and $r_c(0)$ with the length of the dumbbell shows that the 
shape of each of the $A$, $r_c^\star$, $T$ vs $r_c(0)$ curves is independent of $\ell$.
In addition, the minimum values of $A$ and $r_c^\star$ grow linearly with $\ell$, whereas the minimum value of $T$ is proportional to $\ell^2$.

\begin{figure}[t!]
\begin{center}
\captionsetup[subfigure]{labelformat=empty,labelsep=none}
\subfloat[]
{\includegraphics[scale = 0.4, trim = 0cm 0cm 0cm 0cm, clip]{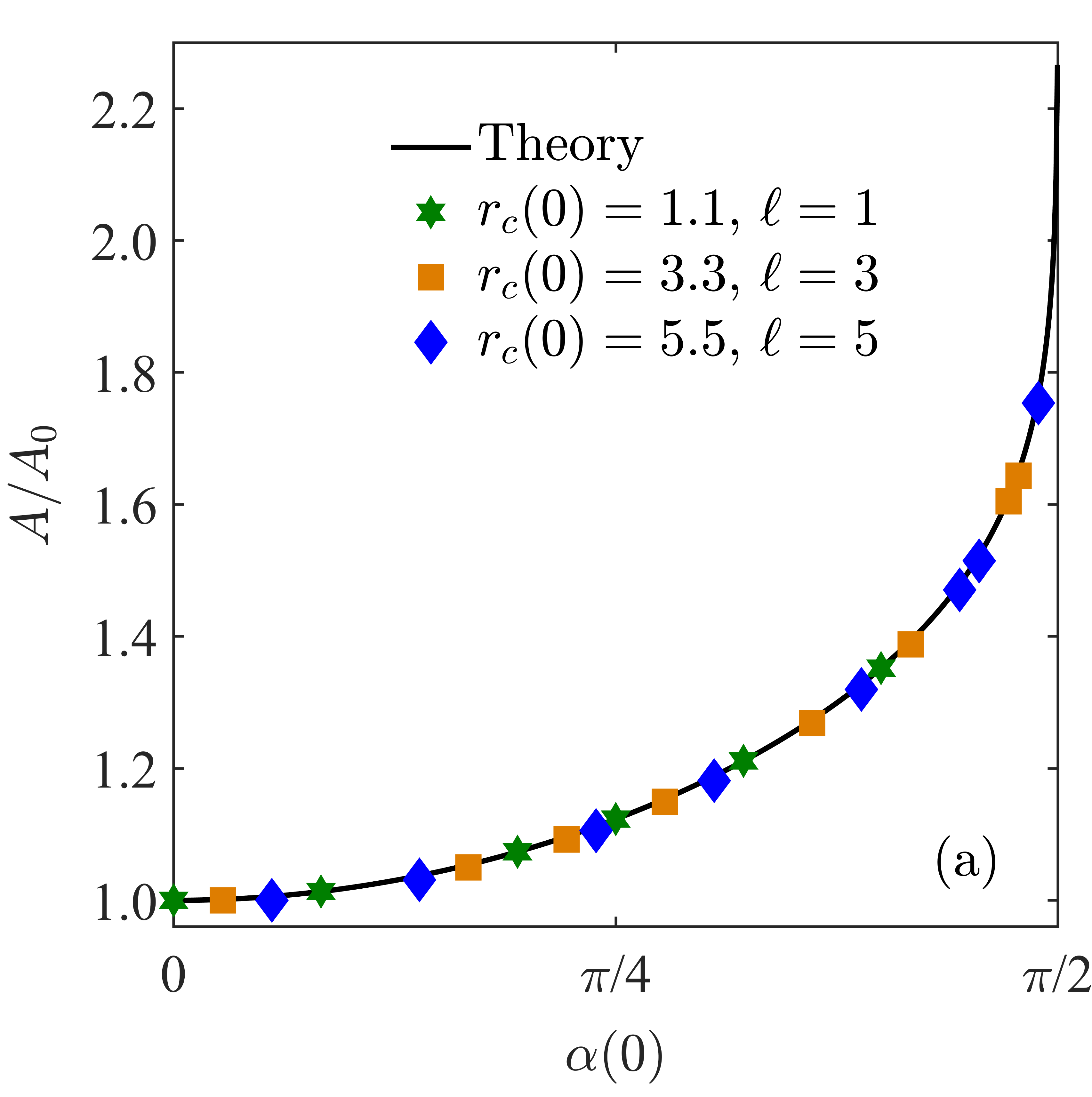}}
\subfloat[]
{\includegraphics[scale = 0.4, trim = 0cm 0cm 0cm 0cm,clip]{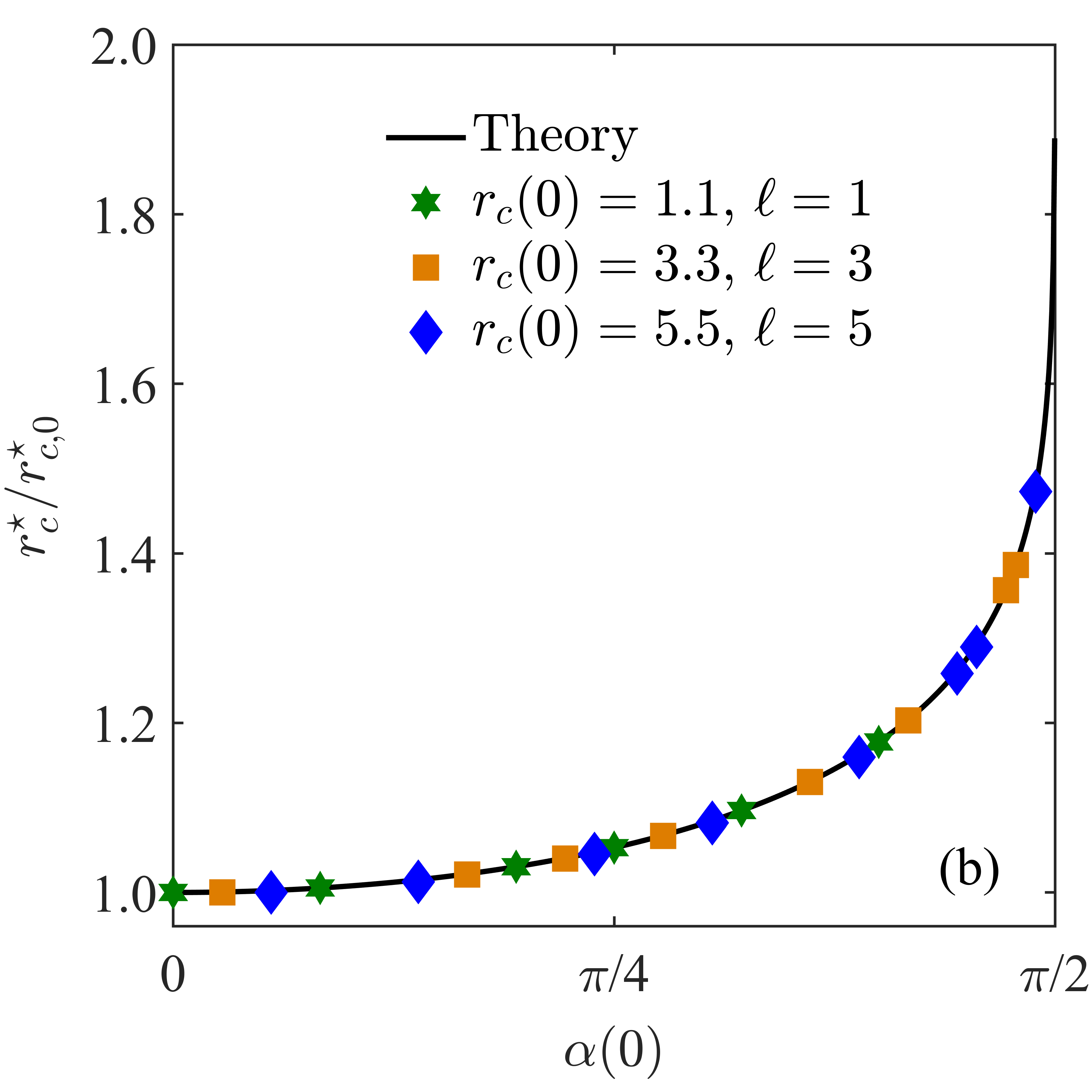}}
\subfloat[]
{\includegraphics[scale = 0.4, trim = 0cm 0cm 0cm 0cm, clip]{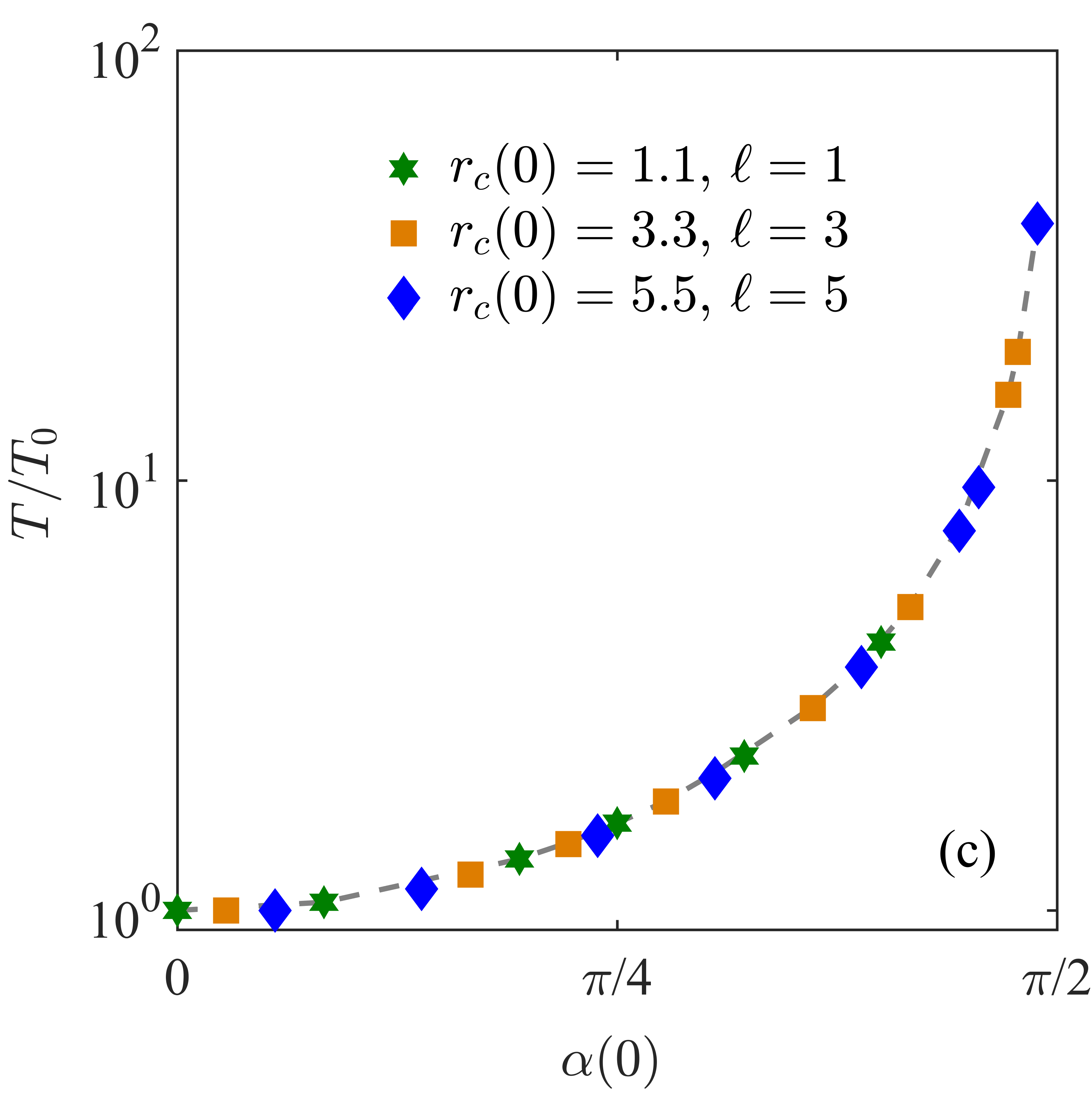}}
\caption{Motion near a Lamb--Oseen vortex: dependence on the initial orientation $\alpha(0)$ of (a) the amplitude $A$, (b) the distance around which the center of mass oscillates, $r_c^{\star}$, and (c) the time period $T$ for $r_c(0)/\ell=1.1$.
$A_0$, $r_{c,0}^*$, $T_0$ are the values of  $A$, $r_{c}^*$, $T$ at $r_c(0)/\ell=1.1$ and $\alpha(0)=0$.
The solid lines are obtained from Eq.~\eqref{eq:A+rstar} (see Sect.~\ref{sect:analytical}); the dashed line is included to guide the eye.
}
\label{Artvsalpha}
\end{center}
\end{figure}

For a fixed $r_c(0)\neq \ell/2$ and different values of $\ell$, the dependence of $A$, $r_c^\star$, and $T$ on the initial orientation of the
dumbbell is shown in Fig.~\ref{Artvsalpha}. Only the range $0\leqslant\alpha(0)<\pi/2$ is shown, since the curves for other ranges 
of $\alpha(0)$ can be obtained by symmetry arguments.
The oscillations are narrow when the dumbbell is initially oriented along the radial direction $\alpha(0)=0$ 
and become wider and wider as the initial orientation approaches the direction tangential to the streamlines of the vortex [$\alpha(0)=\pi/2$].
Once again, the behaviour of the $A$, $r_c^{\star}$, $T$ vs $\alpha(0)$ curves is independent of $r_c(0)/\ell$, 
and the curves for different $\ell$ can be overlapped with suitable normalization.

\begin{figure}[t]
	\begin{center}
		\captionsetup[subfigure]{labelformat=empty,labelsep=none}
		\subfloat[]{\includegraphics[scale=0.51]{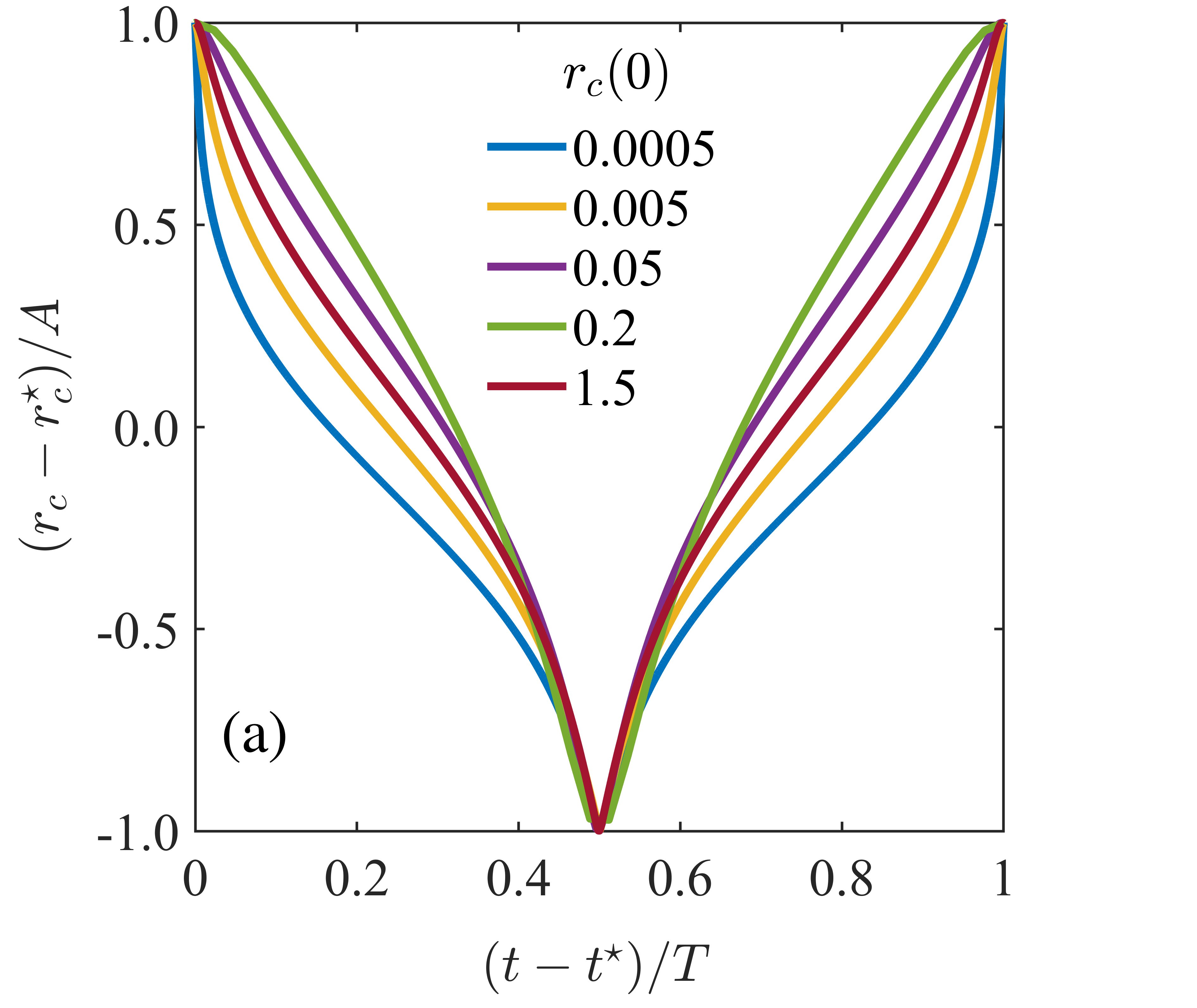}} \hspace{-0.5cm}
		\subfloat[]{\includegraphics[scale=0.55]{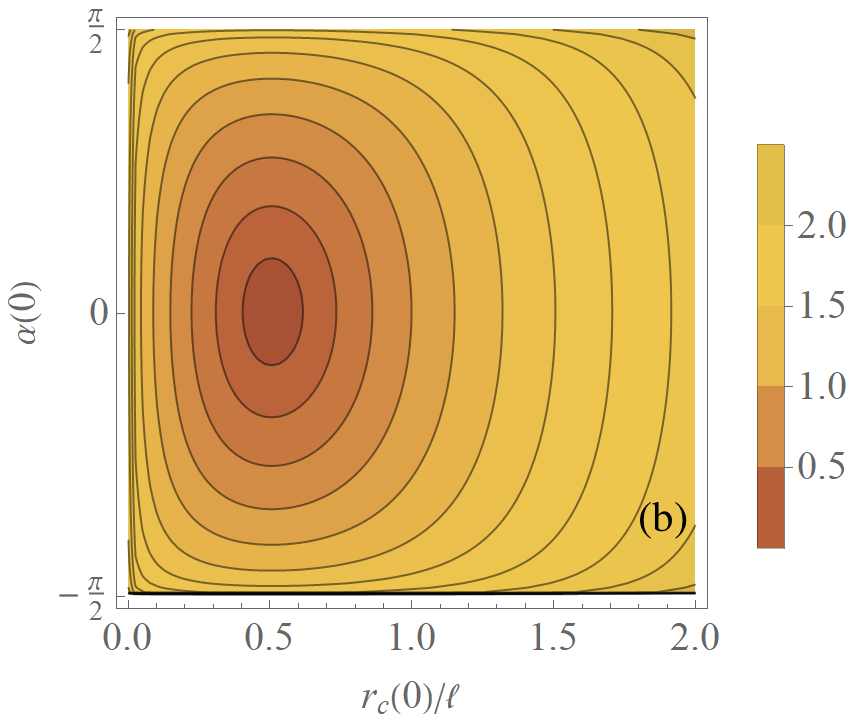}}
		\hspace{0.5mm}
		\caption{(a) Functional shape of the radial oscillation in the Lamb--Oseen vortex for $\alpha(0)=-\pi/4$ and different $r_c(0)$. 
			(b) Contour plot of the amplitude $A$ of the radial oscillation.}
		\label{fig:functionshape}
	\end{center}
\end{figure}

Figure~\ref{fig:functionshape}(a) indicates that not only features such as
the magnitude and the time period, but even the functional shape of the radial oscillation varies with the initial
configuration and the system parameters.

In the next Section, we show that the above numerical observations can be explained by studying of Eqs.~\eqref{eq:cm+ell}.

\section{Dynamics in the $(r_c,\alpha)$ plane}
\label{sect:analytical}
The evolution equations for the variables $r_c$, $\varphi_c$, $\alpha$ can be derived from
Eqs.~\eqref{eq:cm+ell} (see the Appendix) and take the form
\begin{subequations}
\begin{eqnarray}
\label{reqn}
\dot{r}_c & = &
- \frac{\ell \sin{\alpha}}{4}\,[\Omega(r_1)-\Omega(r_2)], 
\\[2mm]
\label{alphaeqn}
\dot{\alpha} & = &
\cos{\alpha}\bigg( \frac{r_c}{\ell} - \frac{\ell}{4r_c} \bigg)[\Omega(r_1)-\Omega(r_2)], 
\\[2mm]
\label{phieqn}
\dot{\varphi}_c & = &
\frac{1}{2}\left\{ \Omega(r_1)+\Omega(r_2) + \frac{\ell\cos{\alpha}}{2r_c}\,[\Omega(r_1)-\Omega(r_2)]\right\}.
\end{eqnarray}
\label{eq:r-alpha-phi}%
\end{subequations}
In the above equations, the distances of the beads from the center of the vortex are expressed in terms of $r_c$ and $\alpha$ as
\begin{equation}
r_1^2 = r_c^2+\frac{\ell^2}{4}+\ell\,r_c\cos{\alpha}, \qquad
r_2^2 = r_c^2+\frac{\ell^2}{4}-\ell\,r_c\cos{\alpha},
\label{eq:distances}
\end{equation}
which follows from 
\begin{equation}
\label{eq:r1&r2}
\bm r_1=\bm r_c+\frac{\bm\ell}{2}, \qquad \bm r_2=\bm r_c-\frac{\bm\ell}{2}.
\end{equation}
An immediate consequence of Eqs.~\eqref{eq:r-alpha-phi} is that,
for a linear velocity field [$U(r)\propto r$], the dumbbell performs a solid-body rotation at fixed distance and orientation
[the same conclusion could also be reached by noting that the tension in the connector vanishes for a linear velocity 
field---see Eq.~\eqref{eq:tension}---and the beads move as tracers].
In the following analysis, therefore, it will be assumed
that the velocity depends on the radial distance in a nonlinear way.

Furthermore, the right-hand sides of Eqs.~\eqref{reqn} and~\eqref{alphaeqn} do not depend on the polar angle $\varphi_c$. Hence
$\varphi_c$ is ``slaved'' to the variables $r_c$ and $\alpha$, and the main features of the dynamics can be
understood by focusing on the $(r_c,\alpha)$ plane alone. In particular, the Poincar\'e--Bendixson theorem implies that 
the motion cannot be chaotic \cite{ccv10}.

In the $(r_c,\alpha)$ plane, the system possesses the following sets of fixed points, each of which corresponds to a solid-body rotation of the 
dumbbell in physical space:
\begin{itemize}
\item $\mathcal{P}_1=\{(\ell/2,0),(\ell/2,\pi)\}$. In these two configurations,
one of the beads stays at the vortex center, while the other rotates on a circle of radius $\ell$, so that the dumbbell rotates around
one of its ends;
\item $\mathcal{P}_2=\{(r_c,\alpha) \text{ s.t. } r_c=0\}$.
The center of mass stays at the vortex center and the dumbbell rotates on itself with the beads moving on the circle of radius $\ell/2$.
As a matter of fact, the existence of
this fixed point cannot be deduced from Eqs.~\eqref{eq:r-alpha-phi}, because neither $\alpha$ nor $\varphi_c$ are defined when $r_c=0$.
However, it follows directly from Eq.~\eqref{eq:cm+ell}, since $\bm u(\bm x_1)=-\bm u(\bm x_2)$ when $r_c=0$;
\item $\mathcal{P}_3=\{(r_c,\alpha) \text{ s.t. $r_c>0$ and $\alpha=\pm\pi/2$}\}$.
Both the beads rotate with the flow on the same circle of radius $r_1=r_2$,
and the dumbbell moves tangentially to the circle of radius $r_c$;
\item $\mathcal{P}_4=\{(r_c,\alpha) \text{ s.t. $r_c>0$, $\alpha\neq\pm\pi/2$, $\Omega(r_1)=\Omega(r_2)$}\}$.
The dumbbell rotates at a fixed distance from the vortex centre
while keeping its orientation with respect to the radial direction.
Note that these fixed points only exist if $\Omega(r)$ goes through the same value at two or more different radial locations of $r$.
\end{itemize}
It can be checked that in all the above cases the radial velocity of the center of mass is zero. In addition,
the beads experience no tension and move with the flow as fluid particles, \textit{i.e.} $\dot{\bm x}_i=\bm u(\bm x_i)$, $i=1,2$.
This can be seen by using Eqs.~\eqref{velocity} and~\eqref{eq:tension} and noting that
\begin{itemize}
\item For the two points in $\mathcal{P}_1$, we have either
$\bm u(\bm x_1)=0$ and $\bm\ell\perp\bm u(\bm x_2)$ or $\bm u(\bm x_2)=0$ and $\bm\ell\perp\bm u(\bm x_2)$;  
\item For the point in $\mathcal{P}_2$, the connector $\bm\ell$ is perpendicular to both $\bm u(\bm x_1)$ and $\bm u(\bm x_2)$; 
\item The configurations belonging to the sets
$\mathcal{P}_3$ and  $\mathcal{P}_4$ satisfy $\bm u(\bm r_1)\cdot{\bm\ell}=-U(r_1)\,r_2\hat{\bm\varphi}_1\cdot\hat{\bm r}_2=
U(r_2)\,r_1\hat{\bm \varphi}_2 \cdot\hat{\bm r}_1=\bm u(\bm r_2)\cdot{\bm\ell}$.
\end{itemize}
From the analysis below, it will be clear that the fixed points in $\mathcal{P}_2$ and $\mathcal{P}_3$ are unstable, whereas 
those in $\mathcal{P}_1$ are neutrally 
stable. The nature of the points belonging to $\mathcal{P}_4$, when they exist, depends on the form of the fluid angular velocity.
Obviously, the fixed points  of the $(r_c,\alpha)$ plane become periodic orbits in the $(r_c,\varphi_c,\alpha)$ space which correspond to 
a solid-body rotation of the dumbbell at a constant angular velocity [Eq.~\eqref{phieqn} indeed
yields $\varphi_c(t)=\varphi_c(0)+\overline{\Omega}t$ with $\overline{\Omega}=\Omega(r_1)$ or $\overline{\Omega}=\Omega(r_2)$].

The points $\mathcal{P}_3$ impact the dynamics of the dumbbell in the same way for any vortex flow. These points, indeed,
form two straight lines ($\alpha=\pm\pi/2$) which separate the 
domain into two disconnected regions, so that the dynamics takes place in either of the stripes $-\pi/2<\alpha<\pi/2$ or
$\pi/2<\alpha<3\pi/2$ depending on the initial orientation of the dumbbell [Fig.~\ref{fig:vectorplot}(a)]. 
As a consequence, the dumbbell never reverts its orientation
with respect to the radial direction and the sign of $\cos\alpha$ remains constant during the time evolution, as was observed in
Sect.~\ref{sect:spirographic} in the case of the Lamb--Oseen vortex. 

Finally, a very general result can be deduced from Eqs.~\eqref{reqn} and~\eqref{alphaeqn}.
These equations indeed display the same dependence on $\Omega(r_1)$ and $\Omega(r_2)$
and can be combined to yield
\begin{equation}
\dfrac{d\alpha}{dr_c/\ell}=-{4} \bigg( \frac{r_c}{\ell} - \frac{\ell}{4r_c} \bigg)\cot\alpha.
\end{equation}
It follows that
\begin{equation}
\frac{r_c}{\ell}\, e^{-2(r_c/\ell)^2} \cos\alpha = \text{constant}
\label{eq:traj}
\end{equation}
is a constant of motion for all vortices. The implications of this result for the dynamics of the dumbbell depend on how the fluid angular velocity behaves as a function of $r$. 

\subsection{Decreasing fluid angular velocity}
\label{sect:decreasing}

\begin{figure}[t]
\begin{center}
\captionsetup[subfigure]{labelformat=empty,labelsep=none}
\hspace{-0.6cm}
\subfloat[]
{\includegraphics[width=0.32\textwidth]{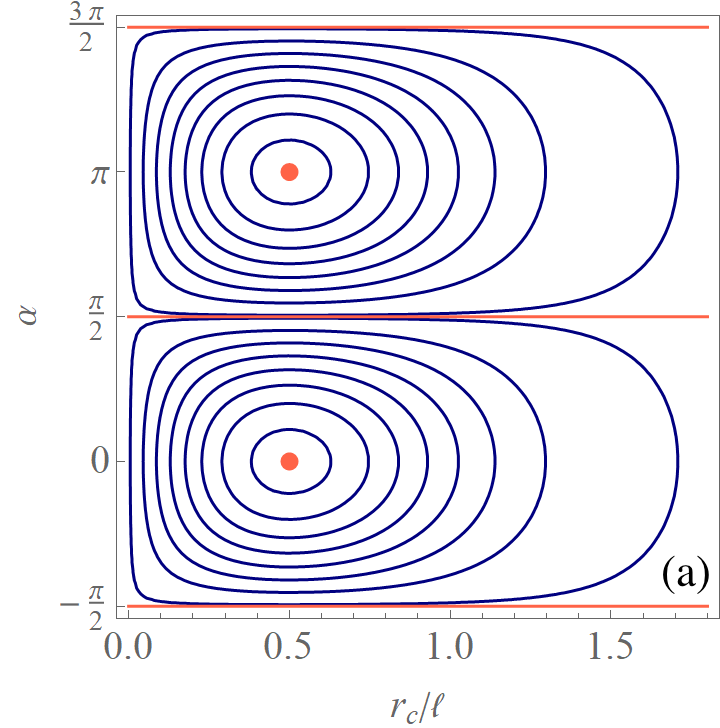}}
\hspace{0mm}
\subfloat[]
{\includegraphics[width=0.37\textwidth]{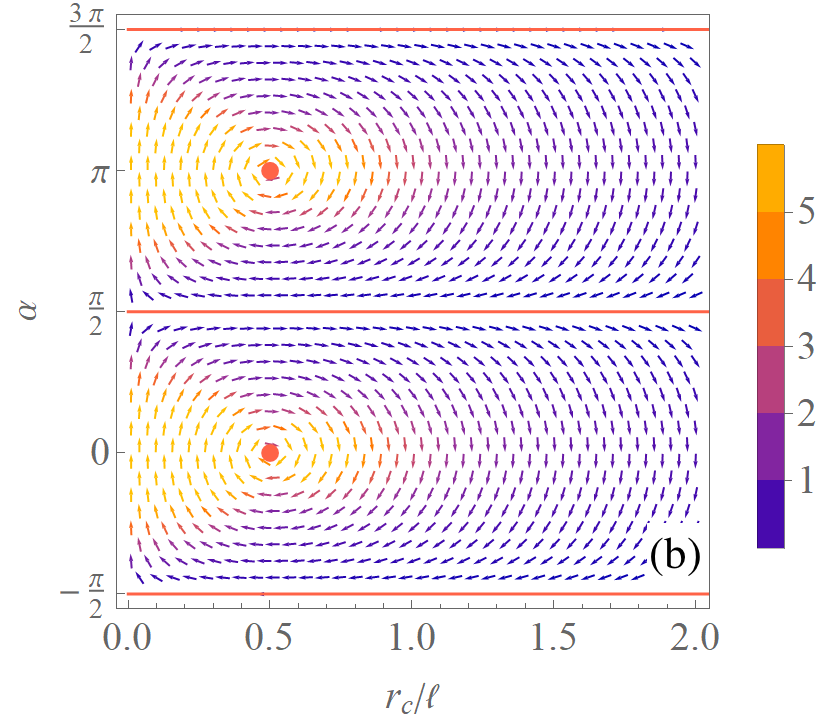}}
\hspace{0.5mm}
\subfloat[]
{\includegraphics[width=0.32\textwidth]{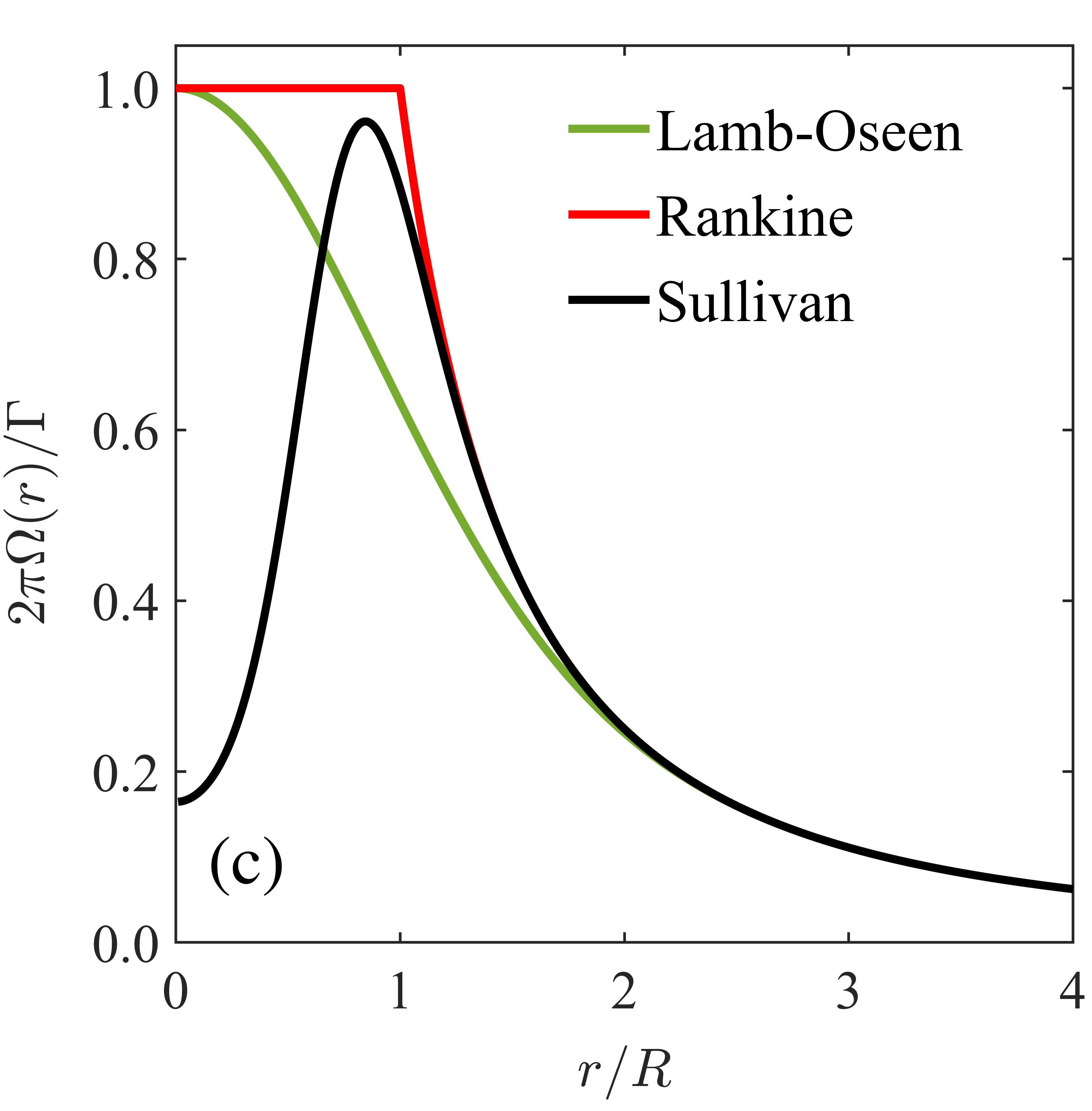}}
\caption{(a) Fixed points (red) and periodic orbits (blue) in the $(r_c,\alpha)$ plane for a Lamb--Oseen vortex with $R=0.1$ and $\ell=1$. 
The red points are the set $\mathcal{P}_1$ and the red lines are the set $\mathcal{P}_3$.
(b) Vector plot of the field $(\dot{r}_c,\dot{\alpha})$ for a Lamb--Oseen vortex with $R=0.1$ and $\ell=1$.  
The colour of the arrows is proportional to the magnitude of the 
vector $(\dot{r}_c,\dot{\alpha})$. (c) Profiles of the fluid angular
velocity for the Lamb--Oseen (green), Rankine (red), and Sullivan (black)
vortices.}
\label{fig:vectorplot}
\end{center}
\end{figure}

A wide class of single vortices, which includes the Lamb--Oseen vortex, the point vortex, and axisymmetric vortices with $\Omega(r)\propto 1/r^{p}$ 
($0\leqslant p\leqslant 2$) \cite{rg15},
has angular vorticity $\Omega(r)$ decreasing with increase in $r$. We recall that in this case the set $\mathcal{P}_4$ is empty.
For such vortices, 
Eq.~\eqref{eq:traj} indicates that the trajectories in the $(r_c,\alpha)$ plane form a family of closed orbits 
around the fixed points $(\ell/2,0)$ and $(\ell/2,\pi)$ [see Fig.~\ref{fig:vectorplot}(a)].  
Therefore, the variables $r_c$ and $\alpha$ are periodic functions of time with same period.
The orbits are parametrized by the initial conditions $r_c(0)$ and $\alpha(0)$.

The radial oscillation of the center of mass reverses its direction (inward or ouwards) when $\alpha=0,\pi$, that is when the connector
is parallel or antiparallel to the radial direction. Hence, for a given orbit, the minimum and maximum values of $r_c$, denoted as $r_{\rm min}$ and $r_{\max}$, 
are the two roots of the equation
\begin{equation}
\label{eq:A+rstar}
\frac{r_c}{\ell} \, e^{-2(r_c/\ell)^2} =  \frac{r_c(0)}{\ell}\, e^{-2[r_c(0)/\ell]^2} \vert\cos\alpha(0)\vert.
\end{equation}
By using Eq.~\eqref{eq:A+rstar}, it is thus possible
to calculate the amplitude and the distance around which the oscillation takes place as $A=(r_{\rm max}-r_{\rm min})/2$ and
$r_c^\star=(r_{\rm max}+r_{\rm min})/2$, respectively. The solid lines in Figs.~\ref{Artvsr0}(a,b) and~\ref{Artvsalpha}(a,b)
and the contour plot of $A$ in Fig.~\ref{fig:functionshape}(b) have been obtained in this way.
Since $(r_c/\ell) e^{-2r_c^2/\ell^2}$ is a concave function of $r_c$ and vanishes as $r_c$ tends to either zero or infinity,
both $A$ and $r_c^\star$ diverge when either $\alpha(0)$ approaches $\pm\pi/2$ or $r_c(0)$ tends to zero or infinity.
For such initial configurations, the centre of mass performs very wide oscillations, as was noted in Sect.~\ref{sect:spirographic}.
Moreover, the solutions of Eq.~\eqref{eq:A+rstar} do not depend on $r_c$ and $\ell$ separately, but only on the ratio $r_c/\ell$. 
Hence the functional dependence of $A$ and $r_c^\star$ on $r_c(0)$ and
$\alpha(0)$ is independent of $\ell$ and, for fixed $r_c(0)$ and $\alpha(0)$, the values of $A$ and $r_c^\star$ are proportional to $\ell$
(see Figs.~\ref{Artvsr0} and~\ref{Artvsalpha}). Figure~\ref{fig:functionshape}(b) also shows that the dynamics becomes less and less
sensitive to the initial orientation as $r_c(0)$ is increased.

The correlation between the orientation of the dumbbell and the direction of its radial motion, shown in Fig.~\ref{timeseries}(b), can also be predicted from Eq.~\eqref{reqn}.
Indeed, if $\Omega(r)$ is decreasing, then the sign of $\Omega(r_1)-\Omega(r_2)$ is fixed at the beginning of the evolution
(recall that during the motion the dumbbell never reverses its orientation with respect to the radial direction).
Therefore, whether 
the radial motion is inward or outward is entirely determined by the sign of $\sin\alpha(t)$.

It ought be stressed that Eq.~\eqref{eq:traj} is independent of $\Omega(r)$. Therefore, all the properties of the dynamics that have been mentioned
so far are independent of the form of the vortex, provided that $\Omega(r)$ decreases with increasing $r$.
In particular, the dependence of $A$ and $r_c^\star$ on the initial configuration of the dumbbell 
[see the solid lines in Figs.~\ref{Artvsr0}(a,b) and~\ref{Artvsalpha}(a,b) and the contour plot of $A$ in Fig.~\ref{fig:functionshape}] is the same irrespective of the functional form of $\Omega(r)$.
What varies with the specific form of the vortex is the speed at which the orbits in the $(r_c,\alpha)$ plane are covered, which in turn
determines the evolution of the angle $\varphi_c$ and ultimately the shape of the spirographic trajectories
in physical space.
Therefore, the behavior of $T$ which was shown in Figs.~\ref{Artvsr0}(c) and~\ref{Artvsalpha}(c)
is not generic but is specific to the Lamb--Oseen vortex.
To explain this further, in Fig.~\ref{fig:vectorplot}(b) we show a vector plot of the field $(\dot{r}_c,\dot{\alpha})$ for the Lamb--Oseen vortex
where the color of the arrows is a function of the magnitude of the vector field. Clearly, the orbits of the system are those 
described in Fig.~\ref{fig:vectorplot}(a), which are the same for any vortex with decreasing $\Omega(r)$. However, the speed of the system along such orbits depends on the details of the Lamb--Oseen vortex. 
A different vortex would perform the same orbits but at a different speed. It would thus generate spirographic trajectories with same
amplitude and at same radial distance, but of a different shape.

Finally, since the evolution of $r_c$ and $\alpha$ is periodic, the right-hand side of Eq.~\eqref{phieqn} is also periodic with same time period $T$. As a consequence, the evolution of $\varphi_c$ can be written as
\begin{equation}
    \varphi_c(t)=\varphi_c(0)+\omega t + \Phi(t),
\end{equation}
where $\Phi(t)$ is a periodic function of period $T$ and $\omega$ is the average of the right-hand side of Eq.~\eqref{phieqn} over a time period. In general, 
$2\pi/\omega$ differs from $T$, and therefore the rotational motion is not periodic.
This explains the behavior observed
in Sect.~\ref{sect:spirographic}, where the time evolution of $\varphi_c$ was found to be the combination of a linear growth and a periodic oscillation of period $T$ superposed to it [see Figs.~\ref{timeseries}(d) and (e)].

\subsection{Rankine vortex}

It was mentioned above that for the set $\mathcal{P}_4$ to be non-empty, the fluid angular velocity must be a non-monotonic function of the radial distance.
To explore how this additional set of fixed points may modify the dynamics of the dumbbell,
we thus consider vortices such that $\Omega(r)$ is not strictly decreasing.
We start with the Rankine vortex \cite{saffman,wmz06}, whose spatial structure is simple enough to allow an analytical study. The Rankine vortex
indeed consists of an inner core of size $R$ which is in  solid-body rotation  and an outer region where the flow is potential {[Fig.~\ref{fig:vectorplot}(c)]}:
\begin{equation}
\Omega(r)=\begin{cases}
\dfrac{\Gamma }{2\pi R^2}, & r\leqslant R,
\\[4mm]
\dfrac{\Gamma}{2\pi r^2}, & r> R.
\end{cases}
\end{equation}
%
%
Compared to vortices with decreasing angular velocity,
there exists a new set of fixed points in the $(r_c,\alpha)$ plane. This
corresponds to configurations in which both the beads lie in the solid-body-rotation core:
\begin{align}
  \mathcal{P}_4=&\left\{(r_c,\alpha) \text{ s.t. } \text{$-\pi/2<\alpha<\pi/2$ and 
$r_1^2=r_c^2+\frac{\ell^2}{4} + \ell\,r_c\cos{\alpha}\leqslant R^2$}\right\}
\\&\cup
\left\{(r_c,\alpha) \text{ s.t. } \text{$\pi/2<\alpha<3\pi/2$ and 
$r_2^2=r_c^2+\frac{\ell^2}{4} - \ell\,r_c\cos{\alpha}\leqslant R^2$}\right\}.
  \label{eq:sbr}
\end{align}
The interior of $\mathcal{P}_4$ obviously is neutrally stable. In contrast, a linear stability analysis shows that the boundary 
of $\mathcal{P}_4$ is stable for $\sin\alpha < 0$ and unstable
for $\sin\alpha >0$. The unstable (stable) portions of the boundary act as a repelling (attracting) set for
the trajectories that start outside $\mathcal{P}_4$ (see Fig.~\ref{fig:rankinesullivanvectorplots}).  

Three different regimes can be identified depending on the ratio $\ell/R$
[see the vector plots of the field $(\dot{r}_c,\dot{\alpha})$ 
in Fig.~\ref{fig:rankinesullivanvectorplots}]: 
\begin{figure}[t]	
	\begin{minipage}[c]{.5\textwidth}
		\includegraphics[width=0.9\textwidth]{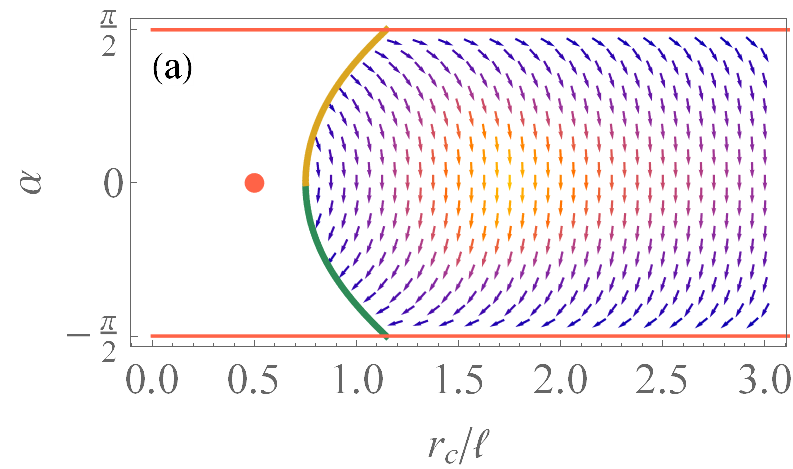}
		\vspace{1mm}
		\includegraphics[width=0.9\textwidth]{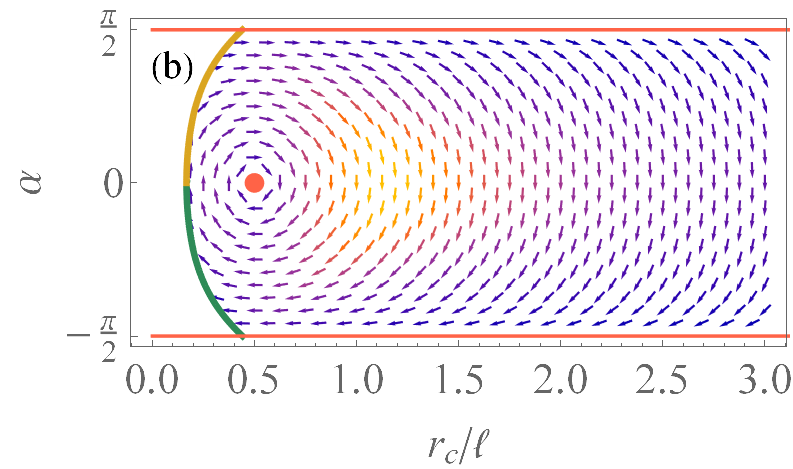}
		\vspace{1mm}
		\end{minipage}%
   \begin{minipage}[c]{.5\textwidth}
	\includegraphics[width=0.9\textwidth]{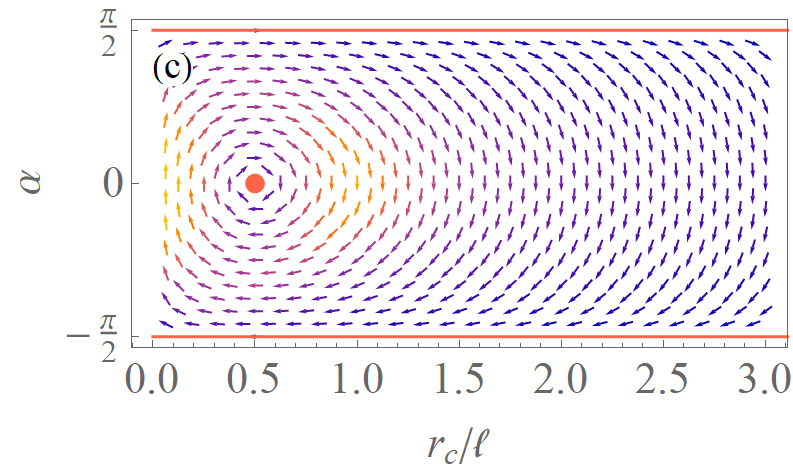}
	\vspace{1mm}
	\includegraphics[width=0.9\textwidth]{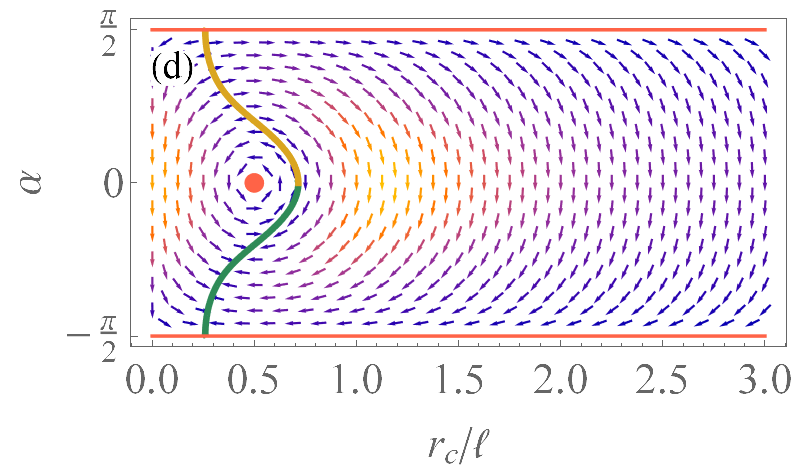}
   \end{minipage}%
	\caption{Vector plots of the field $(\dot{r}_c,\dot{\alpha})$ for a Rankine vortex with (a) $\ell/R=0.8$, (b) $\ell/R=1.5$, (c) $\ell/R=2$. The white area is the interior of $\mathcal{P}_4$ and corresponds to those initial configurations for which the dumbbell is in solid body rotation from the beginning. 
	The green and orange lines are the stable and unstable boundaries of $\mathcal{P}_4$, respectively. (d) Vector plot of the field $(\dot{r}_c,\dot{\alpha})$ for a Sullivan vortex with $\ell/R=1.5$. The orange (green) line is the unstable (stable) subset of $\mathcal{P}_4$. In all plots, the red points are $\mathcal{P}_1$ and the red straight lines are $\mathcal{P}_3$, as in Fig.~\ref{fig:vectorplot}. Only the range $-\pi/2\leqslant\alpha\leqslant\pi/2$ is shown, since the vector fields in the range $\pi/2\leqslant\alpha<3\pi/2$ are identical.
	$\Gamma=2\pi$ and $R=1$ in all cases.}
	\label{fig:rankinesullivanvectorplots}
\end{figure}
\begin{enumerate} [label=(\roman*)]
\item  if $0 < \ell \leqslant  R$, the fixed points $(\ell/2,0)$ and $(\ell/2,\pi)$ lie inside $\mathcal{P}_4$ [Fig.~\ref{fig:rankinesullivanvectorplots}(a)]. 
Therefore, if the system 
starts outside $\mathcal{P}_4$ or on its repulsing boundary, it eventually ends up on the attracting boundary of $\mathcal{P}_4$. 
Periodic orbits are not possible in this case: either the dumbbell is in solid-body rotation from the very beginning  or it ends up in solid-body
rotation after a transient. Note that the motion towards the attracting set continues to
take place along the curves described by Eq. \eqref{eq:traj}, even though now the orbits are not performed in full.

\item if $R < \ell \leqslant  2R$, the fixed points $(\ell/2,0)$ and $(\ell/2,\pi)$ lie outside $\mathcal{P}_4$ [Fig.~\ref{fig:rankinesullivanvectorplots}(b)].
Periodic orbits  are now possible for
initial conditions close to $(\ell/2,0)$ and $(\ell/2,\pi)$.
These periodic orbits are given by Eq. \eqref{eq:traj} and are therefore the same as for any vortex with decreasing angular velocity. What varies
is the speed at which the orbits are performed.

\item if $\ell > 2R$, the set $\mathcal{P}_4$ is empty [see Fig.~\ref{fig:rankinesullivanvectorplots}(c)]. 
The dumbbell is indeed too long
compared to $R$  for both the beads to lie inside the solid-body-rotation core. In this case, the dynamics is qualitatively similar to
that described in Sect.\ref{sect:decreasing} and consists of periodic orbits around either $(\ell/2,0)$ or
$(\ell/2,\pi)$ depending on the value of $\alpha(0)$.
\end{enumerate}

\begin{figure}[t]
	\begin{center}
		\captionsetup[subfigure]{labelformat=empty,labelsep=none}
		\subfloat[]
		{\includegraphics[width=0.33\textwidth]{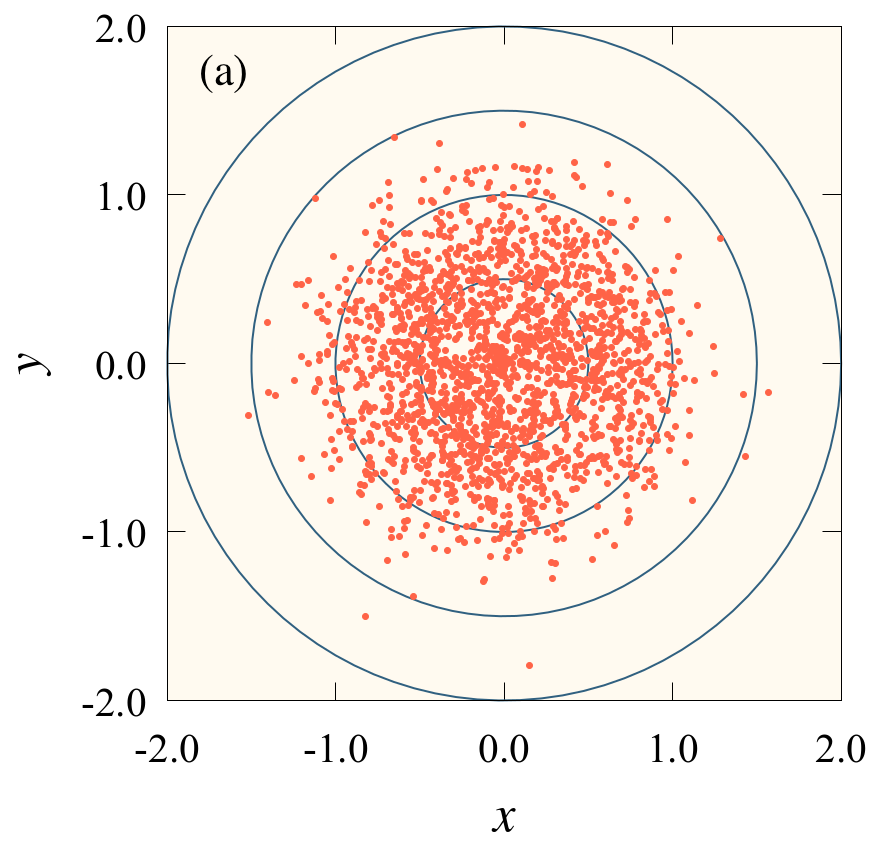}}
		\subfloat[]
		{\includegraphics[width=0.33\textwidth]{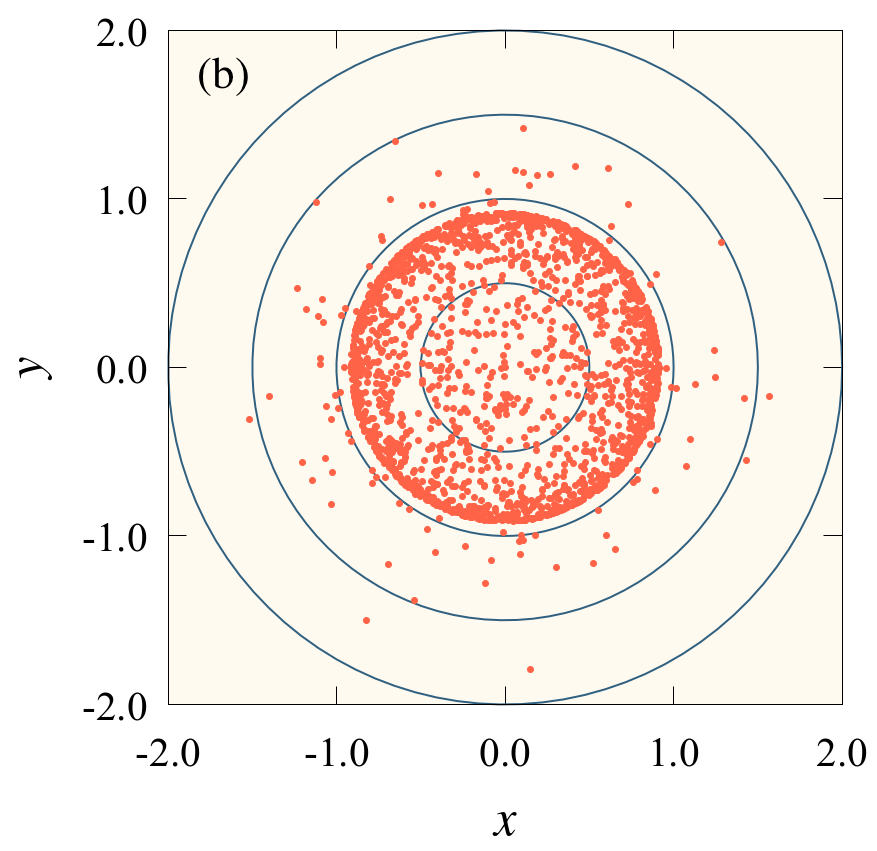}}
		\subfloat[]{\includegraphics[width=0.33\textwidth]{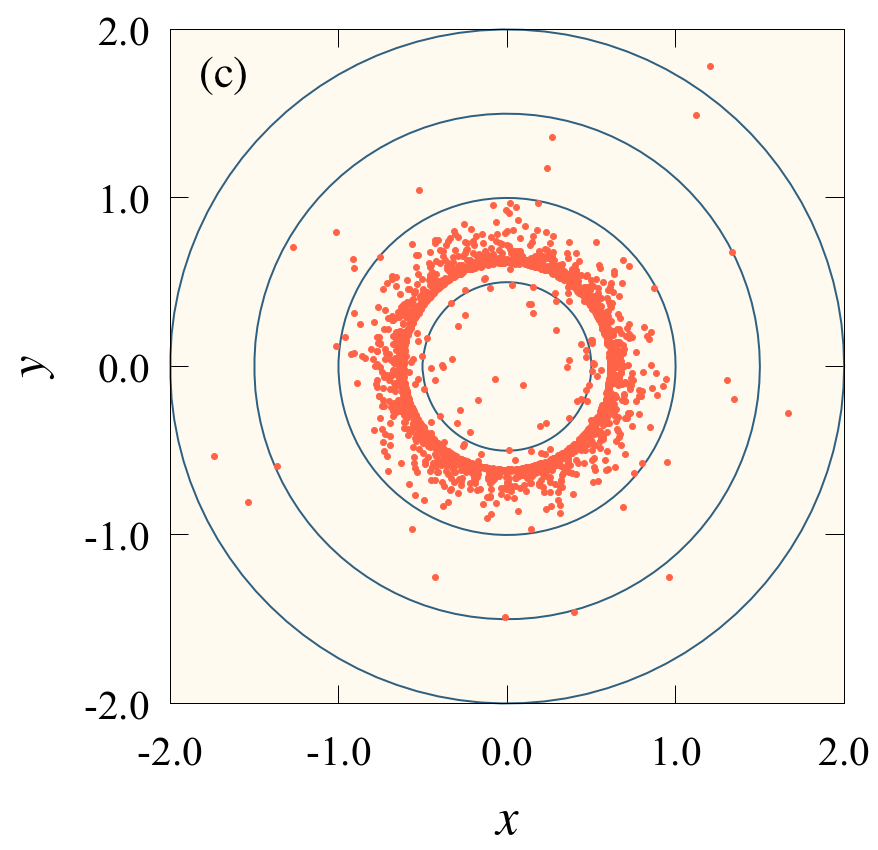}}
		\caption{Spatial distribution of the centers of mass of $2\times 10^3$ non-interacting
dumbbells at $t=200$ in (a) the Lamb--Oseen vortex for $\ell=0.8$, (b) the Rankine vortex for $\ell=0.8$, and (c) the Sullivan vortex for $\ell=1.2$. At $t=0$ the centers of mass
of the dumbbells are distributed uniformly over a disk of radius $r=1.6$ in the Lamb--Oseen and Rankine vortices and $r=1.5$ in the Sullivan vortex. In both the Rankine and Sullivan vortices, $\Gamma=2\pi$ and $R=1$. The parameters of the Lamb--Oseen vortex are the same as in Sect.~\ref{sect:spirographic}.}
		\label{fig:multipledumbbell_lastsnapshots}
	\end{center}
\end{figure}

To show further how the existence of an attracting set modifies the dynamics, in Fig.~\ref{fig:multipledumbbell_lastsnapshots} we compare the long-time 
spatial distribution of an ensembe of dumbbells in the Lamb--Oseen and Rankine vortices (see also Supplemental Movies 3 to 5 \cite{LambOseenmovie,Rankinemovie,AnnulusmovieRankine}).
Naturally, this should only be regarded as a way to visualize the attracting set and not as a realistic simulation of
an ensemble of dumbbells. The latter, indeed, would require accounting for mechanical and hydrodynamic 
interactions between dumbbells, which are instead disregarded here.

In the Lamb--Oseen vortex, the dumbbells spread around the vortex center 
while performing spirographic trajectories with different amplitudes and 
at different distances from the vortex center, and no pattern emerges in their spatial distribution
(see Fig.~\ref{fig:multipledumbbell_lastsnapshots}(a) and Supplemental Movie 3 \cite{LambOseenmovie}). 
In the Rankine vortex, the dynamics is similar to that in the Lamb--Oseen vortex if $\ell>2R$ (not shown).
When $\ell\leqslant 2R$
the dumbbells that start entirely inside the $r\leqslant R$ disk perform a solid-body rotation, while
those that have at least one bead outside the $r\leqslant R$ disk display a different behavior according to their length and initial configuration.
If $0 < \ell \leqslant R$, all such dumbbells 
eventually end up performing a solid-body rotation in the annulus $R - \ell/2 \leqslant r \leqslant \sqrt{R^2 - \ell^2/4}$,
the inner and outer radii of which are determined by the location of the boundary of $\mathcal{P}_4$ at $\alpha=0,\pi$ and
$\alpha=\pm\pi/2,3\pi/2$, respectively.
If $R < \ell \leqslant 2R$, the dumbbells that have an initial configuration which is far from $r_c=\ell/2$, $\alpha=0,\pi$ are attracted inside the
aforementioned annulus, whereas those that start in a configuration close to $r_c=\ell/2$, $\alpha=0,\pi$ 
perform spirographic trajectories. In this case, the long-time spatial distribution of the centers of mass
consists of a core which is in solid-body rotation and an oscillating halo around the distance $r=\ell/2$ (see Fig.~\ref{fig:multipledumbbell_lastsnapshots}(b) and Supplemental Movie 4 \cite{Rankinemovie}).
It is interesting to note that, when $\ell\leqslant 2R$,
the boundary of $\mathcal{P}_4$ in the $(r_c,\alpha)$ plane acts as
a transport barrier that prevents the centers of mass of the dumbbells from penetrating inside the $r< R - \ell/2$ disk from outside. 
Therefore, if the initial distribution of the dumbbells is such that $r_c(0)> R - \ell/2$ for all them,
then the $r< R - \ell/2$ disk remains empty at later times (see Supplemental Movie 5 \cite{AnnulusmovieRankine}).

The study of the Rankine vortex reveals two main differences with the case of decreasing fluid angular velocity.
First, the ratio $\ell/R$ is now an important parameter which discriminates between different dynamical regimes. 
Second, an attracting set emerges, which was absent in vortices with decreasing $\Omega(r)$.
Since the specific shape of this set plays a crucial role, the dynamics of the dumbbell in vortices with non-decreasing angular velocity does not
enjoy the same degree of universality as in the case of a decreasing $\Omega(r)$. To illustrate this further, we consider a two-dimensional
version of the Sullivan vortex. This can no longer be solved analytically but has a smooth angular velocity.

\subsection{Two-dimensional Sullivan vortex}

Sullivan \cite{sullivan1959two} found an exact vortex solution of the three-dimensional Navier--Stokes equations with a  two-cell spatial structure,
\textit{i.e.} with a region of reverse flow near to the axis of the vortex (see also Refs.~\cite{saffman,wmz06}).
The fluid angular velocity $\Omega(r)$ displays a maximum at a given distance from the vortex center {[Fig.~\ref{fig:vectorplot}(c)]}. 
This can be used to construct a stable vortex solution of the two-dimensional Euler equations with non-monotonic angular velocity.
$\Omega(r)$ takes the form
\begin{equation}
\Omega(r) = \frac{\Gamma}{2 \pi r^2}\bigg[ \frac{H(\xi)}{H(\infty)} \bigg], 
\end{equation}
where $\xi = c\, (r/R)^2$ and the function $H(\xi)$ is expressed as
\begin{equation}
H(\xi) = \int_{0}^{\xi} \exp \bigg[ -s+3\int_{0}^{s} \bigg[ \frac{1-e^{-\sigma}}{\sigma} \bigg] d\sigma \bigg] ds.
\end{equation}
The constant $c\approx 6.238$ is chosen in such a way that the maximum of $\Omega(r)$ is at $r\approx R$ \cite{v98}.

In each of the stripes $-\pi/2<\alpha<\pi/2$ and $\pi/2<\alpha<3\pi/2$ of the $(r_c,\alpha)$ plane,
the set of fixed points $\mathcal{P}_4$ forms again a line that divides the stripe into two separate regions [Fig.~\ref{fig:rankinesullivanvectorplots}(d)].
The line consists of an attracting and a repelling portion, and its shape varies with $\ell/R$.
The set $\mathcal{P}_4$ now corresponds to those configurations in which 
one of the beads lies at $r<R$ while the other is at $r>R$ and $\Omega(r_1) = \Omega(r_2)$. 
Two different behaviors can be observed in the $(r_c, \alpha)$ plane [see the vector plot in Fig.~\ref{fig:rankinesullivanvectorplots}(d)].
If the dumbbell starts sufficiently close to ($\ell/2$, 0) or ($\ell/2$, $\pi$), then it performs periodic orbits according to Eq.~\eqref{eq:traj};
otherwise it eventually ends up on the attracting portion of $\mathcal{P}_4$.

To visualize the dynamics and show how it is influenced by the presence of the set $\mathcal{P}_4$, we have again simulated the motion of the center of mass of an ensemble of dumbbells (see Supplemental Movie~6 \cite{Sullivanmovie}). 
For simplicity, in the simulations we have used an approximation of the Sullivan angular-velocity profile that was proposed by Wood and Brown \cite{wb11}:
$\Omega(r)= 0.89 \,r (r/R)^{2.4}[0.3+0.7(r/R)^{7.89}]^{-0.435}$.
The dumbbells whose initial conditions $(r_c(0),\alpha(0))$ are close 
to ($\ell/2, 0$) or ($\ell/2, \pi$) perform spirographic trajectories in an annulus around $r=\ell/2$.
Those that have an initial configuration $(r_c(0),\alpha(0))$ 
far from ($\ell/2,0$) and ($\ell/2, \pi$) with $r_c(0) < \ell/2$ ($r_c(0) > \ell/2$) move away from (move towards) the vortex center and eventually end up 
performing solid-body rotation. Consequently, the long-time spatial distribution of the
dumbbells in the Sullivan vortex consists of an annulus which is in solid-body rotation an oscillating halo around $r=\ell/2$ 
(see Fig.~\ref{fig:multipledumbbell_lastsnapshots}(c) and Supplemental Movie~6 \cite{Sullivanmovie}). 

Thus, the example of the two-dimensional Sullivan vortex further demonstrates that if $\Omega(r)$ does not decrease with $r$,
the attracting set that emerges in the $(r_c,\alpha)$ plane strongly impacts the dynamics of the dumbbell in a way that is specific to the
particular form of the vortex. Different dynamical regimes may in principle be generated by modifying the funcional dependence of
$\Omega(r)$ on $r$.

\section{Summary and concluding remarks}
\label{sect:conclusions}
This study investigates the motion of particles in a vortex flow
by going beyond the point-particle approximation.
It thus aims to be a step in the direction of a better
understanding of the dynamics of extended objects in a flow field.
In the case of a rigid dumbbell, the simplicity of the system allows a
detailed analysis of the motion and of its dependence on the properties
of the vortex.

The main result is that, in the class of two-dimensional steady vortices
with angular velocity decreasing as a function of the radial distance,
the center of mass of a rigid dumbbell performs spirographic
trajectories around the vortex center.
The qualitative features of the dynamics do not depend on the  details of
the vortex. For instance, the amplitude of the radial oscillation
and the distance around which the oscillation is
performed are fully independent of the functional form of the vortex.
The situation changes when the fluid angular velocity is not strictly
monotonic. An attracting set emerges in the configuration space,
and this impacts the dynamics in a way that depends on the details of
the vortex.

The analysis is restricted to steady vortices, but several results also
apply to time-dependent vortices. In particular, the quantity
$(r_c/\ell)\exp(-2r_c^2/\ell^2)\cos\alpha$ remains a constant of motion even
for a time-dependent vortex and
the orbits in the $(r_c,\alpha)$ plane are unchanged:
only
the way these orbits are covered varies according to the temporal
evolution of the fluid angular velocity.
Two-dimensional turbulent flows forced at large scales are characterized
by large long-lived vortices in the vicinity of which straining is weak. A dumbbell would typically
remain in a given vorticity-dominated region for a long time, during which the quantity
$(r_c/\ell)\exp(-2r_c^2/\ell^2)\cos\alpha$ would remain constant. It would be interesting to explore the
consequences of this conserved quantity for the dynamics of dumbbells in 
two-dimensional turbulence.

The study also disregards Brownian fluctuations. However, an inspection
of the vector plots in Figs.~\ref{fig:vectorplot} and~\ref{fig:rankinesullivanvectorplots} shows that, for most initial
configurations,
Brownian fluctuations would only cause small perturbations of the
spirographic dynamics.
In contrast, inertial effects may have a strong impact. If the inertia
of the beads is not negligible, the dumbbell is likely
to acquire a nonzero mean radial velocity resulting in its ejection or
entrapment depending on the ratio between the bead and fluid density
\cite{rm97}.
Nevertheless, we have seen that the instantaneous radial
velocity of the dumbbell depends on its orientation. It would therefore
be interesting to study whether the orientation dynamics of the
dumbbell speeds up or slows down its ejection or entrapment. 

Finally, in a dumbbell only the two beads interact with the fluid, and
hence the drag force is concentrated at the ends of the object.
Nevertheless, based on the above analysis of the spirographic dynamics,
we expect that a rigid fiber would perform a qualitatively similar
motion, even though the effects of the
hydrodynamic interactions between the segments of the fiber remain to be
understood.

\begin{acknowledgments}
The authors are grateful to Giorgio Krstulovic and Jason R.~Picardo for helpful discussions. SRY acknowledges the financial support from the CNRS
through the 80 $|$ Prime program. RG acknowledges support of the Department of Atomic Energy, Government of India, under project no. RTI4001. DV acknowledges the support of the Indo--French Center for Applied Mathematics (IFCAM). 
\end{acknowledgments}

\appendix*

\section{}\label{phasespaceeqn}

Recall that, in polar coordinates, the orthogonal bases at the positions of the beads and of the centre of mass 
are denoted as $\{\hat{\bm r}_i,\hat{\bm\varphi}_i\}$ $(i=1,2)$
and $\{\hat{\bm r}_c,\hat{\bm\varphi}_c\}$, respectively.
These obey the relationships
\begin{equation}
	\qquad
	\hat{\bm r}_1\cdot\hat{\bm \varphi}_2 = - \hat{\bm r}_2\cdot\hat{\bm \varphi}_1, 
	\qquad
	\hat{\bm\varphi}_c=\frac{r_1}{2r_c}\,\hat{\bm\varphi}_1+\frac{r_2}{2r_c}\,\hat{\bm\varphi}_2
\end{equation}
whence
\begin{equation}
	\bm\ell\cdot\hat{\bm\varphi}_c=\frac{r_1r_2}{r_c}\,\hat{\bm r}_1\cdot\hat{\bm\varphi}_2=
	-\frac{r_1r_2}{r_c}\,\hat{\bm r}_2\cdot\hat{\bm\varphi}_1.
	\label{eq:lphic}
\end{equation}
In addition
\begin{equation}
	\bm{\ell} \cdot \hat{\bm r}_c = \ell \cos{\alpha}, 
	\qquad
	\bm{\ell} \cdot \hat{\bm{\varphi}}_c = \ell \sin{\alpha}.
	\label{eq:lalpha}
\end{equation}
By using the definition of the velocity in Eqs.~\eqref{eq:velocity} and~\eqref{eq:omega} as well as
Eq.~\eqref{eq:lphic} and the second of Eqs.~\eqref{eq:lalpha}, we find
\begin{equation}
	\bm u(\bm r_1)\cdot\bm r_2 = -r_c\,\ell\,\Omega(r_1)\sin\alpha,
	\qquad
	\bm u(\bm r_2)\cdot\bm r_1 = r_c\,\ell\,\Omega(r_2)\sin\alpha.
\end{equation}
Thus, Eq.~\eqref{eq:cm} yields
\begin{equation}\label{eq:radialvel}
	\dot{r}_c = \frac{1}{2}[\bm{u}(\bm{r}_1,t)+\bm{u}(\bm{r}_2,t)] \cdot \hat{\bm r}_c
	= \frac{1}{4r_c}[\bm u(\bm r_1)\cdot\bm r_2 + \bm u(\bm r_2)\cdot\bm r_1]
	= -\frac{\ell \sin{\alpha}}{4} [\Omega(r_1)-\Omega(r_2)],
\end{equation}
which is Eq.~\eqref{reqn}.
To derive the evolution equation for $\alpha$, we first note that
\begin{equation}\label{eq:cos}
	r_c\,\frac{d \cos{\alpha}}{dt}=\frac{1}{\ell}\frac{d }{dt}(\bm{\ell} \cdot \bm{r}_c)-\dot{r}_c\,\cos\alpha.
\end{equation}
Then, Eq.~\eqref{eq:ell} yields
\begin{equation}\label{eq:ldotrc}
	\frac{d}{dt}(\bm{\ell} \cdot \bm{r}_c) = \bm{\ell} \cdot \dot{\bm r}_c + \bm{r}_c \cdot \dot{\bm \ell}
	= -\{\hat{\bm{\ell}} \cdot [\bm{u}(\bm{r}_1,t)- \bm{u}(\bm{r}_2,t)]\} \,  (\hat{\bm{\ell}}\cdot{\bm r}_c)
	=-r_c^2\sin\alpha\cos\alpha\, [\Omega(r_1)-\Omega(r_2)]
\end{equation}
By using Eqs.~\eqref{eq:radialvel} and \eqref{eq:ldotrc} in Eq.~\eqref{eq:cos}, we find
\begin{equation}
	\frac{d \cos{\alpha}}{dt} = -\sin\alpha\cos{\alpha}\bigg( \frac{r_c}{\ell} - \frac{\ell}{4r_c} \bigg)[\Omega(r_1)-\Omega(r_2)], 
\end{equation}
which gives Eq.~\eqref{alphaeqn}.
The $x$-component of Eq.~\eqref{eq:cm} may now be used to derive an evolution equation for $\varphi_c$:
\begin{equation}\label{eq:phi1}
	r_c\, \frac{d \cos{\varphi_c}}{dt} = \hat{\bm x}\cdot\dot{\bm r}_c -  \dot{r}_c\cos{\varphi_c}=
	\hat{\bm x}\cdot[\dot{\bm r}_c -  \dot{r}_c \hat{\bm r}_c].
\end{equation}
Note that Eq.~\eqref{eq:r1&r2} implies 
\begin{equation}
	\label{eq:ellperp}
	r_1\hat{\bm\varphi}_1 = r_c\hat{\bm\varphi}_c +\bm\ell^\perp/2, \qquad
	r_2\hat{\bm\varphi}_2 = r_c\hat{\bm\varphi}_c -\bm\ell^\perp/2,
\end{equation}
where 
\begin{equation}
	\bm \ell^\perp=-\ell\sin\alpha\,\hat{\bm r}_c+\ell\cos\alpha\,\hat{\bm\varphi}_c
\end{equation}
is such that $\bm\ell\cdot\bm\ell^\perp=0$.
By using Eqs.~\eqref{eq:ellperp}, we can rewrite Eq.~\eqref{eq:cm} as
\begin{equation}
	\dot{\bm r}_c = \frac{r_c}{2}[\Omega(r_1)+\Omega(r_2)]\hat{\bm\varphi}_c
	+\frac{\bm\ell^\perp}{4}[\Omega(r_1)-\Omega(r_2)].
\end{equation}
We thus find
\begin{equation}
	\dot{\bm r}_c -  \dot{r}_c \hat{\bm r}_c = \frac{r_c}{2}[\Omega(r_1)+\Omega(r_2)]\hat{\bm\varphi}_c
	+\frac{\ell}{4}\cos\alpha[\Omega(r_1)-\Omega(r_2)]\hat{\bm\varphi}_c.
\end{equation}
Finally, inserting the latter expression in Eq.~\eqref{eq:phi1} yields
\begin{equation}\label{eq:phi2}
	\frac{d \cos{\varphi_c}}{dt} =-\frac{\sin{\varphi_c}}{2} [\Omega(r_1) + \Omega(r_2)] 
	-\frac{\ell}{4r_c}\cos\alpha\sin{\varphi_c}[\Omega(r_1) - \Omega(r_2)]
\end{equation}
and hence Eq.~\eqref{phieqn}.

\bibliography{refs-dumbbell}

\begin{thebibliography}{27}%
\makeatletter
\providecommand \@ifxundefined [1]{%
 \@ifx{#1\undefined}
}%
\providecommand \@ifnum [1]{%
 \ifnum #1\expandafter \@firstoftwo
 \else \expandafter \@secondoftwo
 \fi
}%
\providecommand \@ifx [1]{%
 \ifx #1\expandafter \@firstoftwo
 \else \expandafter \@secondoftwo
 \fi
}%
\providecommand \natexlab [1]{#1}%
\providecommand \enquote  [1]{``#1''}%
\providecommand \bibnamefont  [1]{#1}%
\providecommand \bibfnamefont [1]{#1}%
\providecommand \citenamefont [1]{#1}%
\providecommand \href@noop [0]{\@secondoftwo}%
\providecommand \href [0]{\begingroup \@sanitize@url \@href}%
\providecommand \@href[1]{\@@startlink{#1}\@@href}%
\providecommand \@@href[1]{\endgroup#1\@@endlink}%
\providecommand \@sanitize@url [0]{\catcode `\\12\catcode `\$12\catcode
  `\&12\catcode `\#12\catcode `\^12\catcode `\_12\catcode `\%12\relax}%
\providecommand \@@startlink[1]{}%
\providecommand \@@endlink[0]{}%
\providecommand \url  [0]{\begingroup\@sanitize@url \@url }%
\providecommand \@url [1]{\endgroup\@href {#1}{\urlprefix }}%
\providecommand \urlprefix  [0]{URL }%
\providecommand \Eprint [0]{\href }%
\providecommand \doibase [0]{https://doi.org/}%
\providecommand \selectlanguage [0]{\@gobble}%
\providecommand \bibinfo  [0]{\@secondoftwo}%
\providecommand \bibfield  [0]{\@secondoftwo}%
\providecommand \translation [1]{[#1]}%
\providecommand \BibitemOpen [0]{}%
\providecommand \bibitemStop [0]{}%
\providecommand \bibitemNoStop [0]{.\EOS\space}%
\providecommand \EOS [0]{\spacefactor3000\relax}%
\providecommand \BibitemShut  [1]{\csname bibitem#1\endcsname}%
\let\auto@bib@innerbib\@empty
\bibitem [{\citenamefont {L\'{a}zaro}\ and\ \citenamefont
  {Lasheras}(1989)}]{ll89}%
  \BibitemOpen
  \bibfield  {author} {\bibinfo {author} {\bibfnamefont {B.~J.}\ \bibnamefont
  {L\'{a}zaro}}\ and\ \bibinfo {author} {\bibfnamefont {J.~C.}\ \bibnamefont
  {Lasheras}},\ }\bibfield  {title} {\bibinfo {title} {Particle dispersion in a
  turbulent, plane, free shear layer},\ }\href@noop {} {\bibfield  {journal}
  {\bibinfo  {journal} {Phys. Fluids A}\ }\textbf {\bibinfo {volume} {1}},\
  \bibinfo {pages} {1035} (\bibinfo {year} {1989})}\BibitemShut {NoStop}%
\bibitem [{\citenamefont {Raju}\ and\ \citenamefont {Meiburg}(1997)}]{rm97}%
  \BibitemOpen
  \bibfield  {author} {\bibinfo {author} {\bibfnamefont {N.}~\bibnamefont
  {Raju}}\ and\ \bibinfo {author} {\bibfnamefont {E.}~\bibnamefont {Meiburg}},\
  }\bibfield  {title} {\bibinfo {title} {Dynamics of small, spherical particles
  in vortical and stagnation point flow fields},\ }\href@noop {} {\bibfield
  {journal} {\bibinfo  {journal} {Phys. Fluids}\ }\textbf {\bibinfo {volume}
  {9}},\ \bibinfo {pages} {299} (\bibinfo {year} {1997})}\BibitemShut {NoStop}%
\bibitem [{\citenamefont {Candelier}\ \emph {et~al.}(2004)\citenamefont
  {Candelier}, \citenamefont {Angilella}, ,\ and\ \citenamefont
  {Souhar}}]{cas04}%
  \BibitemOpen
  \bibfield  {author} {\bibinfo {author} {\bibfnamefont {F.}~\bibnamefont
  {Candelier}}, \bibinfo {author} {\bibfnamefont {J.~R.}\ \bibnamefont
  {Angilella}}, ,\ and\ \bibinfo {author} {\bibfnamefont {M.}~\bibnamefont
  {Souhar}},\ }\bibfield  {title} {\bibinfo {title} {On the effect of the
  {B}oussinesq--{B}asset force on the radial migration of a {S}tokes particle
  in a vortex},\ }\href@noop {} {\bibfield  {journal} {\bibinfo  {journal}
  {Phys. Fluids}\ }\textbf {\bibinfo {volume} {16}},\ \bibinfo {pages} {1765}
  (\bibinfo {year} {2004})}\BibitemShut {NoStop}%
\bibitem [{\citenamefont {Goater}\ and\ \citenamefont {Lawrence}(2004)}]{gl04}%
  \BibitemOpen
  \bibfield  {author} {\bibinfo {author} {\bibfnamefont {A.}~\bibnamefont
  {Goater}}\ and\ \bibinfo {author} {\bibfnamefont {G.~A.}\ \bibnamefont
  {Lawrence}},\ }\bibfield  {title} {\bibinfo {title} {Dispersion of heavy
  particles in an isolated pancake-like vortex},\ }\href@noop {} {\bibfield
  {journal} {\bibinfo  {journal} {J. Environ. Eng. Sci.}\ }\textbf {\bibinfo
  {volume} {3}},\ \bibinfo {pages} {403} (\bibinfo {year} {2004})}\BibitemShut
  {NoStop}%
\bibitem [{\citenamefont {Ravichandrana}\ and\ \citenamefont
  {Govindarajan}(2015)}]{rg15}%
  \BibitemOpen
  \bibfield  {author} {\bibinfo {author} {\bibfnamefont {S.}~\bibnamefont
  {Ravichandrana}}\ and\ \bibinfo {author} {\bibfnamefont {R.}~\bibnamefont
  {Govindarajan}},\ }\bibfield  {title} {\bibinfo {title} {Caustics and
  clustering in the vicinity of a vortex},\ }\href@noop {} {\bibfield
  {journal} {\bibinfo  {journal} {Phys. Fluids}\ }\textbf {\bibinfo {volume}
  {27}},\ \bibinfo {pages} {033305} (\bibinfo {year} {2015})}\BibitemShut
  {NoStop}%
\bibitem [{\citenamefont {Deepu}\ \emph {et~al.}(2017)\citenamefont {Deepu},
  \citenamefont {Ravichandran},\ and\ \citenamefont {Govindarajan}}]{drg17}%
  \BibitemOpen
  \bibfield  {author} {\bibinfo {author} {\bibfnamefont {P.}~\bibnamefont
  {Deepu}}, \bibinfo {author} {\bibfnamefont {S.}~\bibnamefont
  {Ravichandran}},\ and\ \bibinfo {author} {\bibfnamefont {R.}~\bibnamefont
  {Govindarajan}},\ }\bibfield  {title} {\bibinfo {title} {Caustics-induced
  coalescence of small droplets near a vortex},\ }\href@noop {} {\bibfield
  {journal} {\bibinfo  {journal} {Phys. Rev. Fluids}\ }\textbf {\bibinfo
  {volume} {2}},\ \bibinfo {pages} {024305} (\bibinfo {year}
  {2017})}\BibitemShut {NoStop}%
\bibitem [{\citenamefont {Ruetsch}\ and\ \citenamefont {Meiburg}(1993)}]{rm93}%
  \BibitemOpen
  \bibfield  {author} {\bibinfo {author} {\bibfnamefont {G.~R.}\ \bibnamefont
  {Ruetsch}}\ and\ \bibinfo {author} {\bibfnamefont {E.}~\bibnamefont
  {Meiburg}},\ }\bibfield  {title} {\bibinfo {title} {On the motion of small
  spherical bubbles in two-dimensional vertical flows},\ }\href@noop {}
  {\bibfield  {journal} {\bibinfo  {journal} {Phys. Fluids A}\ }\textbf
  {\bibinfo {volume} {5}},\ \bibinfo {pages} {2326} (\bibinfo {year}
  {1993})}\BibitemShut {NoStop}%
\bibitem [{\citenamefont {Sokolov}\ and\ \citenamefont {Aranson}(2016)}]{sa16}%
  \BibitemOpen
  \bibfield  {author} {\bibinfo {author} {\bibfnamefont {A.}~\bibnamefont
  {Sokolov}}\ and\ \bibinfo {author} {\bibfnamefont {I.~S.}\ \bibnamefont
  {Aranson}},\ }\bibfield  {title} {\bibinfo {title} {Rapid expulsion of
  microswimmers by a vortical flow},\ }\href@noop {} {\bibfield  {journal}
  {\bibinfo  {journal} {Nat. Commun.}\ }\textbf {\bibinfo {volume} {7}},\
  \bibinfo {pages} {11114} (\bibinfo {year} {2016})}\BibitemShut {NoStop}%
\bibitem [{\citenamefont {Tarama}\ \emph {et~al.}(2014)\citenamefont {Tarama},
  \citenamefont {Menzel},\ and\ \citenamefont {L\"owen}}]{tml14}%
  \BibitemOpen
  \bibfield  {author} {\bibinfo {author} {\bibfnamefont {M.}~\bibnamefont
  {Tarama}}, \bibinfo {author} {\bibfnamefont {A.~M.}\ \bibnamefont {Menzel}},\
  and\ \bibinfo {author} {\bibfnamefont {H.}~\bibnamefont {L\"owen}},\
  }\bibfield  {title} {\bibinfo {title} {Deformable microswimmer in a swirl:
  Capturing and scattering dynamics},\ }\href@noop {} {\bibfield  {journal}
  {\bibinfo  {journal} {Phys. Rev. E}\ }\textbf {\bibinfo {volume} {90}},\
  \bibinfo {pages} {032907} (\bibinfo {year} {2014})}\BibitemShut {NoStop}%
\bibitem [{\citenamefont {Arguedas-Leiva}\ and\ \citenamefont
  {Wilczek}(2020)}]{wilczek}%
  \BibitemOpen
  \bibfield  {author} {\bibinfo {author} {\bibfnamefont {J.-A.}\ \bibnamefont
  {Arguedas-Leiva}}\ and\ \bibinfo {author} {\bibfnamefont {M.}~\bibnamefont
  {Wilczek}},\ }\bibfield  {title} {\bibinfo {title} {Microswimmers in an
  axisymmetric vortex flow},\ }\href@noop {} {\bibfield  {journal} {\bibinfo
  {journal} {New J. Phys.}\ }\textbf {\bibinfo {volume} {22}},\ \bibinfo
  {pages} {053051} (\bibinfo {year} {2020})}\BibitemShut {NoStop}%
\bibitem [{\citenamefont {Berman}\ \emph {et~al.}(2021)\citenamefont {Berman},
  \citenamefont {Buggeln}, \citenamefont {Brantley}, \citenamefont {Mitchell},\
  and\ \citenamefont {Solomon}}]{bbbms21}%
  \BibitemOpen
  \bibfield  {author} {\bibinfo {author} {\bibfnamefont {S.~A.}\ \bibnamefont
  {Berman}}, \bibinfo {author} {\bibfnamefont {J.}~\bibnamefont {Buggeln}},
  \bibinfo {author} {\bibfnamefont {D.~A.}\ \bibnamefont {Brantley}}, \bibinfo
  {author} {\bibfnamefont {K.~A.}\ \bibnamefont {Mitchell}},\ and\ \bibinfo
  {author} {\bibfnamefont {T.~H.}\ \bibnamefont {Solomon}},\ }\bibfield
  {title} {\bibinfo {title} {Transport barriers to self-propelled particles in
  fluid flows},\ }\href@noop {} {\bibfield  {journal} {\bibinfo  {journal}
  {Phys. Rev. Fluids}\ }\textbf {\bibinfo {volume} {6}},\ \bibinfo {pages}
  {L012501} (\bibinfo {year} {2021})}\BibitemShut {NoStop}%
\bibitem [{\citenamefont {Guedda}\ \emph {et~al.}(2021)\citenamefont {Guedda},
  \citenamefont {Chaiboub}, \citenamefont {Benlahsen},\ and\ \citenamefont
  {Misbah}}]{gcbm21}%
  \BibitemOpen
  \bibfield  {author} {\bibinfo {author} {\bibfnamefont {M.}~\bibnamefont
  {Guedda}}, \bibinfo {author} {\bibfnamefont {J.}~\bibnamefont {Chaiboub}},
  \bibinfo {author} {\bibfnamefont {M.}~\bibnamefont {Benlahsen}},\ and\
  \bibinfo {author} {\bibfnamefont {C.}~\bibnamefont {Misbah}},\ }\bibfield
  {title} {\bibinfo {title} {Exact trajectory solutions of a spherical
  microswimmer under flow and external fields},\ }\href@noop {} {\bibfield
  {journal} {\bibinfo  {journal} {Phys. Rev. Fluids}\ }\textbf {\bibinfo
  {volume} {6}},\ \bibinfo {pages} {074102} (\bibinfo {year}
  {2021})}\BibitemShut {NoStop}%
\bibitem [{\citenamefont {Graham}(2018)}]{graham}%
  \BibitemOpen
  \bibfield  {author} {\bibinfo {author} {\bibfnamefont {M.~D.}\ \bibnamefont
  {Graham}},\ }\href@noop {} {\emph {\bibinfo {title} {Microhydrodynamics,
  Brownian Motion, and Complex Fluids}}}\ (\bibinfo  {publisher} {Cambridge
  University Press},\ \bibinfo {address} {Cambridge, UK},\ \bibinfo {year}
  {2018})\BibitemShut {NoStop}%
\bibitem [{\citenamefont {Besant}(1890)}]{besant}%
  \BibitemOpen
  \bibfield  {author} {\bibinfo {author} {\bibfnamefont {W.~H.}\ \bibnamefont
  {Besant}},\ }\href@noop {} {\emph {\bibinfo {title} {Notes on Roulettes and
  Glissettes}}},\ \bibinfo {edition} {2nd}\ ed.\ (\bibinfo  {publisher}
  {Deighton, Bell \& Co.},\ \bibinfo {address} {Cambridge, England},\ \bibinfo
  {year} {1890})\BibitemShut {NoStop}%
\bibitem [{\citenamefont {Piva}\ and\ \citenamefont {Martino}(2009)}]{pm09}%
  \BibitemOpen
  \bibfield  {author} {\bibinfo {author} {\bibfnamefont {M.~F.}\ \bibnamefont
  {Piva}}\ and\ \bibinfo {author} {\bibfnamefont {G.~R.}\ \bibnamefont
  {Martino}},\ }\bibfield  {title} {\bibinfo {title} {A rigid dumbbell settling
  under gravity in a periodic flow field},\ }\href@noop {} {\bibfield
  {journal} {\bibinfo  {journal} {J. Phys. A: Math. Theor.}\ }\textbf {\bibinfo
  {volume} {42}},\ \bibinfo {pages} {025501} (\bibinfo {year}
  {2009})}\BibitemShut {NoStop}%
\bibitem [{fig()}]{fig1b}%
  \BibitemOpen
  \href@noop {} {}\bibinfo {note} {See \url{https://youtu.be/zswLMOvTaqY} for
  the movie showing the dynamics of a dumbbell of length $\ell=1$ for
  $r_c(0)=0.3$ and $\alpha(0)=0$ in the Lamb--Oseen vortex with $R=0.1$ and
  $\Gamma=2\pi$. The beads in the movie are coloured differently only for the
  visualization purposes.}\BibitemShut {Stop}%
\bibitem [{mov()}]{movie1c}%
  \BibitemOpen
  \href@noop {} {}\bibinfo {note} {See \url{https://youtu.be/IGt0qc5_Js0} for
  the movie showing the dynamics of a dumbbell of length $\ell=1$ for
  $r_c(0)=1.0$ and $\alpha(0)=-\pi/4$ in the Lamb--Oseen vortex with $R=0.1$
  and $\Gamma=2\pi$. The beads in the movie are coloured differently only for
  the visualization purposes.}\BibitemShut {Stop}%
\bibitem [{\citenamefont {Cencini}\ \emph {et~al.}(2009)\citenamefont
  {Cencini}, \citenamefont {Cecconi},\ and\ \citenamefont {Vulpiani}}]{ccv10}%
  \BibitemOpen
  \bibfield  {author} {\bibinfo {author} {\bibfnamefont {M.}~\bibnamefont
  {Cencini}}, \bibinfo {author} {\bibfnamefont {F.}~\bibnamefont {Cecconi}},\
  and\ \bibinfo {author} {\bibfnamefont {A.}~\bibnamefont {Vulpiani}},\
  }\href@noop {} {\emph {\bibinfo {title} {Chaos: From Simple Models to Complex
  Systems}}}\ (\bibinfo  {publisher} {World Scientific},\ \bibinfo {address}
  {Singapore},\ \bibinfo {year} {2009})\BibitemShut {NoStop}%
\bibitem [{\citenamefont {Saffman}(1992)}]{saffman}%
  \BibitemOpen
  \bibfield  {author} {\bibinfo {author} {\bibfnamefont {P.~G.}\ \bibnamefont
  {Saffman}},\ }\href@noop {} {\emph {\bibinfo {title} {Vortex Dynamics}}}\
  (\bibinfo  {publisher} {Cambridge University Press},\ \bibinfo {address}
  {Cambridge, UK},\ \bibinfo {year} {1992})\BibitemShut {NoStop}%
\bibitem [{\citenamefont {Wu}\ \emph {et~al.}(2006)\citenamefont {Wu},
  \citenamefont {Ma},\ and\ \citenamefont {Zhou}}]{wmz06}%
  \BibitemOpen
  \bibfield  {author} {\bibinfo {author} {\bibfnamefont {J.-Z.}\ \bibnamefont
  {Wu}}, \bibinfo {author} {\bibfnamefont {H.-Y.}\ \bibnamefont {Ma}},\ and\
  \bibinfo {author} {\bibfnamefont {M.-D.}\ \bibnamefont {Zhou}},\ }\href@noop
  {} {\emph {\bibinfo {title} {Vorticity and Vortex Dynamics}}}\ (\bibinfo
  {publisher} {Springer-Verlag},\ \bibinfo {address} {Berlin Heidelberg},\
  \bibinfo {year} {2006})\BibitemShut {NoStop}%
\bibitem [{Lam()}]{LambOseenmovie}%
  \BibitemOpen
  \href@noop {} {}\bibinfo {note} {See \url{https://youtu.be/UClfE_Ec25k} for
  the movie showing the dynamics of the centers of mass of $2\times 10^3$
  non-interacting dumbbells of length $\ell=0.8$ in the Lamb--Oseen vortex with
  $R=0.1$ and $\Gamma=2\pi$. At $t=0$, the centers of mass are distributed
  uniformly over a disc of radius 1.6.}\BibitemShut {Stop}%
\bibitem [{Ran()}]{Rankinemovie}%
  \BibitemOpen
  \href@noop {} {}\bibinfo {note} {See
  \url{https://www.youtube.com/watch?v=_8LtHk58E_g&t=10s} for the movie showing
  the dynamics of the centers of mass of $2\times 10^3$ non-interacting
  dumbbells of length $\ell=0.8$ in the Rankine vortex with $R=1$ and
  $\Gamma=2\pi$. At $t=0$, the centers of mass are distributed uniformly over a
  disc of radius 1.6.}\BibitemShut {Stop}%
\bibitem [{Ann()}]{AnnulusmovieRankine}%
  \BibitemOpen
  \href@noop {} {}\bibinfo {note} {See \url{https://youtu.be/G9xDsvV6vcA} for
  the movie showing the dynamics of the centers of mass of $2\times 10^3$
  non-interacting dumbbells of length $\ell=0.8$ in the Rankine vortex with
  $R=1$ and $\Gamma=2\pi$. At $t=0$, the centers of mass are distributed
  uniformly over an annulus of inner and outer radii 0.6 and 1.6,
  respectively.}\BibitemShut {Stop}%
\bibitem [{\citenamefont {Sullivan}(1959)}]{sullivan1959two}%
  \BibitemOpen
  \bibfield  {author} {\bibinfo {author} {\bibfnamefont {R.~D.}\ \bibnamefont
  {Sullivan}},\ }\bibfield  {title} {\bibinfo {title} {A two-cell vortex
  solution of the {N}avier--{S}tokes equations},\ }\href@noop {} {\bibfield
  {journal} {\bibinfo  {journal} {J. Aerosp. Sci.}\ }\textbf {\bibinfo {volume}
  {26}},\ \bibinfo {pages} {767} (\bibinfo {year} {1959})}\BibitemShut
  {NoStop}%
\bibitem [{\citenamefont {Vatistas}(1998)}]{v98}%
  \BibitemOpen
  \bibfield  {author} {\bibinfo {author} {\bibfnamefont {G.~H.}\ \bibnamefont
  {Vatistas}},\ }\bibfield  {title} {\bibinfo {title} {New model for intense
  self-similar vortices},\ }\href@noop {} {\bibfield  {journal} {\bibinfo
  {journal} {J. Prop. Power}\ }\textbf {\bibinfo {volume} {14}},\ \bibinfo
  {pages} {462} (\bibinfo {year} {1998})}\BibitemShut {NoStop}%
\bibitem [{Sul()}]{Sullivanmovie}%
  \BibitemOpen
  \href@noop {} {}\bibinfo {note} {See
  \url{https://www.youtube.com/watch?v=KXHC9wUPvXc} for the movie showing the
  dynamics of the centers of mass of $2\times 10^3$ non-interacting dumbbells
  of length $\ell=1.2$ in the Sullivan vortex with $R=1$ and $\Gamma=2\pi$. At
  $t=0$, the centers of mass are distributed uniformly over a disc of radius
  1.5.}\BibitemShut {Stop}%
\bibitem [{\citenamefont {Wood}\ and\ \citenamefont {Brown}(2011)}]{wb11}%
  \BibitemOpen
  \bibfield  {author} {\bibinfo {author} {\bibfnamefont {V.~T.}\ \bibnamefont
  {Wood}}\ and\ \bibinfo {author} {\bibfnamefont {R.~A.}\ \bibnamefont
  {Brown}},\ }\bibfield  {title} {\bibinfo {title} {Simulated tornadic vortex
  signatures of tornado-like vortices having one- and two-celled structures},\
  }\href@noop {} {\bibfield  {journal} {\bibinfo  {journal} {J. Appl. Meteorol.
  Climatol.}\ }\textbf {\bibinfo {volume} {50}},\ \bibinfo {pages} {2338}
  (\bibinfo {year} {2011})}\BibitemShut {NoStop}%
\end{thebibliography}%

\end{document}